\documentclass[prd,preprint,superscriptaddress,preprintnumbers,eqsecnum,showpacs,nofootinbib,nobibnotes]{revtex4}
\usepackage{amsfonts,bm}
\usepackage{graphicx}
\usepackage{pstricks}

\newcommand{\be}{\begin{equation}}
\newcommand{\bea}{\begin{eqnarray}}
\newcommand{\ee}{\end{equation}}
\newcommand{\eea}{\end{eqnarray}}
\newcommand{\bpi}{\begin{picture}}
\newcommand{\bce}{\begin{center}}
\newcommand{\epi}{\end{picture}}
\newcommand{\ece}{\end{center}}

\def\s#1{{\scriptscriptstyle #1}}

\def\gtree{\Gamma^{(0)}}
\def\gtreeb{\widetilde{\Gamma}^{(0)}}
\def\gfullb{\widetilde{\Gamma}}
\def\bqq{\mathrm{I}\!\Gamma}

\def\bcj{J}
\def\bcjb{\widetilde{\bcj}}

\def\diff{{\rm d}}

\def\NV{\bqq'}
\def\NVindex{\bqq'^{\alpha\mu\nu}}
\def\NP{ V}
\def\mrgi{\overline m}

\begin{document}

\title{The dynamical equation of the effective gluon mass}

\author{A.~C. Aguilar}
\email{Arlene.Aguilar@ufabc.edu.br}
\affiliation{Federal University of ABC, CCNH, \\
Rua Santa Ad\'{e}lia 166, CEP 09210-170, Santo Andr\'{e}, Brazil.}

\author{D. Binosi}
\email{binosi@ectstar.eu}
\affiliation{European Centre for Theoretical Studies in Nuclear
Physics and Related Areas (ECT*) and Fondazione Bruno Kessler, \\Villa Tambosi, Strada delle
Tabarelle 286, 
I-38123 Villazzano (TN)  Italy}

\author{J. Papavassiliou}
\email{Joannis.Papavassiliou@uv.es}
\affiliation{\mbox{Department of Theoretical Physics and IFIC,  
University of Valencia}
E-46100, Valencia, Spain}

\begin{abstract}

In this article we derive the integral equation that controls the 
momentum dependence of the effective gluon mass in the Landau gauge.  
This is accomplished by means of a well-defined separation 
of the corresponding ``one-loop dressed'' Schwinger-Dyson equation  
into two distinct contributions, 
one associated with the mass and one with the standard kinetic 
part of the gluon. The entire construction relies on the
existence of a longitudinally coupled vertex of 
nonperturbative origin, which 
enforces gauge invariance  in the presence of a dynamical mass. 
The specific structure of the resulting mass equation,
supplemented by the additional requirement of a positive-definite 
gluon mass,  imposes a rather stringent constraint on the 
derivative of the gluonic dressing function, which 
is comfortably satisfied by the large-volume lattice data for the gluon propagator,  
both for $SU(2)$ and $SU(3)$.
The numerical treatment of the mass equation, under some simplifying assumptions, 
is presented for the aforementioned gauge groups, giving rise to 
a gluon mass that is a non-monotonic function of the momentum. 
Various theoretical improvements and possible future directions are briefly discussed.

\end{abstract}


\maketitle

\section{Introduction}

A large body of recent 
high-quality lattice results
indicate  that the  gluon
propagator and  the  ghost   dressing  function  of  pure  Yang-Mills
theories,  computed in  the  conventional Landau  gauge, are  infrared (IR)
finite,    both     in    
$SU(2)$~\cite{Cucchieri:2007md,Cucchieri:2007rg,Cucchieri:2009zt,Cucchieri:2011ga,Cucchieri:2011um}    
and    in
$SU(3)$~\cite{Bogolubsky:2007ud,Bowman:2007du,Bogolubsky:2009dc,Oliveira:2009eh}.
These important results have sparked  
a renewed interest in  
the important issue of dynamical mass generation  
in non-Abelian gauge theories, and  especially in QCD~\cite{Aguilar:2006gr,Binosi:2007pi,Aguilar:2008xm,Binosi:2008qk,Binosi:2009qm,Aguilar:2009ke}.
Specifically, 
as has been suggested in a series of works, the 
finiteness of these  quantities 
may be interpreted as a direct consequence of  
the generation of a non-perturbative (momentum-dependent) gluon mass, 
which acts as an IR cutoff of the theory~\cite{Cornwall:1981zr, Aguilar:2007fe}. 
In the picture put forth in these articles, 
the fundamental Lagrangian of the Yang-Mills theory (or that of QCD) is never altered;  
the generation of the gluon mass takes place dynamically, 
without violating any of the underlying symmetries~ \cite{Cornwall:1981zr,Aguilar:2008xm,Binosi:2009qm}.

Given the non-perturbative nature of the mass generating mechanism, 
its study in the continuum proceeds through the Schwinger-Dyson equations (SDEs) 
that govern the dynamics of the various Green's functions 
of the theory~\cite{Binosi:2007pi,Binosi:2008qk,Alkofer:2000wg,Fischer:2006ub,Braun:2007bx}, and especially of the gluon propagator, $\Delta(q^2)$. 
The main conceptual and technical challenge in this context is to obtain 
as a solution of these integral equations 
an IR-finite 
gluon propagator [{\it i.e.}, $\Delta^{-1}(0)= m^2(0)$],  without interfering with the 
gauge invariance (or the BRST symmetry) 
of the theory, encoded in the 
Ward identities (WIs) and Slavnov-Taylor identities (STIs)
satisfied by the Green's functions under study~\cite{Aguilar:2008xm,Binosi:2007pi,Binosi:2008qk}. 
A self-consistent framework for enforcing the crucial property 
of gauge invariance at the level of the {\it truncated SDEs} 
is provided by the  synthesis of the pinch technique~(PT)~\cite{Cornwall:1981zr,
Cornwall:1989gv,Binosi:2002ft,Binosi:2003rr,Binosi:2009qm}
with the background field method~(BFM)~\cite{Abbott:1980hw}. 


In the presence of a dynamically generated mass, the (inverse) Euclidean gluon propagator
assumes the form $\Delta^{-1}(q^2) =q^2 J(q^2) + m^2(q^2)$,  
where the first term corresponds to the ``kinetic term'', or ``wave function'' contribution, 
whereas the second is the (positive-definite) momentum-dependent mass~\cite{Aguilar:2009ke}.  
However, to date,  
practically all studies attempting to determine the 
IR behavior of the gluon propagator from SDEs 
eventually  boil down to the solution of some integral equation 
involving the entire gluon propagator $\Delta(q^2)$~\cite{Cornwall:1981zr,
Aguilar:2008xm,Binosi:2009qm}, rather than 
its two components,  $J(q^2)$ and  $m^2(q^2)$.
This is to be contrasted to what happens in
the analogous studies of chiral symmetry breaking,  
where one derives a system of two coupled equations, 
one determining the ``wave function'' (``kinetic part'')
of the quark self-energy, and one determining the 
dynamical (constituent) quark mass~\cite{Roberts:1994dr,Aguilar:2010cn}. 
Of course, in the 
case of the quark self-energy the above separation 
of both sides of the corresponding SDE (quark gap equation)
is realized in a direct way,  
due to the distinct 
Dirac properties of the two quantities appearing in it, while 
in the case of the gluon propagator no such straightforward  separation is possible. 
However, a well-defined procedure, first outlined in~\cite{Aguilar:2009ke},
and explained here in more detail, allows for an analogous separation 
even in the case of the gluon propagator. 
The purpose of the present article is to 
identify and isolate 
from the SDE of the (Landau gauge) gluon propagator 
the dynamical equation that determines the 
evolution of the gluon mass, study its main properties, 
and find approximate solutions for  $m^2(q^2)$.


As has been emphasized in some of the literature cited above,  
a crucial condition 
for the realization of the gluon mass generation scenario  
is the existence of a {\it longitudinally coupled vertex},
to be denoted by $V$, which 
must be added to the conventional (fully-dressed) three-gluon vertex, 
denoted by $\bqq$~\cite{Aguilar:2009ke}.
Specifically, the vertex  $\NV = \bqq +V$
satisfies the same STIs as $\bqq$, but now replacing the gluon propagators appearing on their
rhs by a massive ones (schematically, $\Delta \to \Delta_m$).  
The dynamical reason for the emergence of this special vertex, 
as well as its diagrammatic realization in terms of Feynman graphs,   
is intimately connected to the well-known Schwinger mechanism~\cite{Schwinger:1962tn,Schwinger:1962tp},  
which enables the 
non-perturbative generation of a gauge-boson mass. 
In particular, one assumes that the strong QCD dynamics give rise to 
longitudinally-coupled 
composite (bound-state) massless poles~\cite{Jackiw:1973tr,Cornwall:1973ts,Eichten:1974et,Jackiw:1973ha, Poggio:1974qs,Farhi:1982vt,Frampton:2008zz}.
These poles act like Nambu-Goldstone  excitations, in the sense that 
they preserve the form of the STIs of the theory 
in the presence of a mass, but they are not associated with the breaking of any 
local or global symmetry.

It turns out that 
the way the vertex $V$ generates the mass at the level of 
the SDE is by introducing a ``deviation'' from the 
so-called  ``seagull identity'' [given in Eq.~(\ref{seagull})]. 
The role of this identity  is to enforce the masslessness 
of a gauge boson (gluon or photon) when massive propagators 
appear inside its loops, assuming always that the 
WI and STI's are maintained, {\it i.e.}, the transversality of 
the (gluon or photon) self-energy is preserved. 
For example, as explained 
in~\cite{Aguilar:2009ke}, in scalar QED it is exactly this 
identity that enforces the masslessness of the photon 
at the level of the ``one-loop dressed'' SDE; in this case 
the massive propagator entering into the loop is that of the charged scalar field.
The crucial point is that if the ``massive'' STI 
were to be enforced by only modifying $\bqq$
[{\it i.e.}, by carrying out the replacement  $\Delta \to \Delta_m$ 
in the closed expressions obtained for  $\bqq$ 
by solving the STIs it satisfies, see Eq.~(\ref{X10})], 
then the  ``seagull identity'' 
would force the (would-be) gluon mass to vanish, {\it i.e.}, would lead to 
the invalidation of the entire mass generation mechanism. The fact that the 
missing part for satisfying the ``massive'' STI is instead provided by the longitudinally coupled $V$ 
has the far-reaching consequence of finally furnishing a non-trivial
equation for the mass.  
Thus, the equation for the gluon mass is determined  
as {\it the amount by which the seagull cancellation is distorted 
due to the presence of the vertex $V$}.

To be sure, one could in principle determine 
the closed form of $V$ by resorting to a procedure similar to 
that described in~\cite{Binosi:2011wi} for $\bqq$, namely write down the most general 
structure allowed for a longitudinal vertex 
with three Lorentz indices, and then determine the corresponding form factors from 
the WI and STI that this vertex must satisfy. 
It turns out, however, that 
$V$ enters into the SDE for the Landau gauge gluon propagator 
(in the PT-BFM scheme) in a very particular way, 
which renders its closed form unnecessary; 
all one needs for the derivation of the mass equation is 
to postulate the existence of $V$ ({\it i.e.}, assume that it is not identically zero) 
and that it satisfies the required WI and STIs. 

In principle, 
the mass equation obtained in Eq.~(\ref{me-final}) 
must be accompanied by the corresponding equation determining the kinetic term 
$J(q^2)$; the solution of the resulting system of two coupled integral equations 
will then furnish the behavior of $m^2(q^2)$ and $J(q^2)$, and therefore 
that of $\Delta(q^2)$. The technical limitation in realizing these 
procedure is the dependence of the  equation for $J(q^2)$
on the various form factors comprising the ghost-gluon kernel; 
the latter enters into play through the form of the vertex $\bqq$ 
[see Eqs.~(\ref{X10})-(\ref{abcde})] .
The way to circumvent this problem is to actually 
solve Eq.~(\ref{me-final}) for  $m^2(q^2)$ using as input for the $\Delta$
appearing in it the available lattice data, both for $SU(2)$~\cite{Cucchieri:2007md} and $SU(3)$~\cite{Bogolubsky:2007ud}. 

It turns out that the specific form of the mass equation in Eq.~(\ref{me-final})  introduces a 
non-trivial constraint 
on the precise behavior that $\Delta$ must display in the region between \mbox{(1-5) $\rm GeV^2$}. 
Specifically, in order for the gluon mass to be positive definite,
the first derivative 
of the quantity $ q^2 \Delta(q^2)$ (the ``gluon dressing function'')
must furnish a sufficiently {\it negative} 
contribution in the aforementioned range of momenta. Interestingly enough, the 
$\Delta$ obtained from the lattice has indeed this particular property. This is to be contrasted 
to what happens, for example, in the case of a simple massive propagator $1/(q^2+m^2)$  
or with the Gribov-Zwanziger propagator $q^2/(q^4+m^4)$ (with $m$ constant)~\cite{Gribov:1977wm,Zwanziger:1993dh}; 
the derivatives of the corresponding dressing functions, 
$q^2/(q^2+m^2)$ and $q^4/(q^4+m^4)$, respectively,  
are positive in the entire range of (Euclidean) momenta,
thus excluding the possibility of a positive-definite gluon mass.

The article is organized as follows. 
In Section~\ref{gf} we introduce the necessary notation and review briefly 
the aspects of the  PT-BFM formalism relevant to this work.
In Section~\ref{vert} we explain in detail the modifications 
that must be introduced to the three-gluon vertex of the theory in order 
to treat the generation  of a gluon mass in a 
gauge invariant way ({\it i.e.}, preserving the STIs of the theory).  In particular, 
the importance of the nonperturbative vertex $V$ and its special properties are emphasized, 
and the changes introduced to the corresponding SDE during the  
transition from massless to massive solutions are discussed in detail. 
In Section~\ref{st} we outline  the 
precise criteria that will lead to the separation of the 
SDE for the gluon propagator into two equations, one for the kinetic part and one for the mass.
The central role of the ``seagull-identity'' in carrying out this separation is stressed, 
and some explicit characteristic calculations are presented. 
Then, in Section~\ref{gml} we combine the ingredients introduced in the previous sections 
and derive the final form of the dynamical equation for the gluon mass in the Landau gauge. 
In Section~\ref{numan} we first study the implications of the gluon mass equation 
in the limit of vanishing physical momentum. Then 
we solve an approximate form of this equation, using lattice data 
as input for the ``unknown'' quantity $\Delta$.
The solution for the gluon mass so obtained is then appropriately ``subtracted out''
from  $\Delta$, giving rise to an estimate for the quantity $J(q^2)$. 
These ingredients are then combined to construct   
the renormalization-group~(RG) invariant gluon mass, appearing in the usual definition 
of the effective QCD charge within the PT-BFM framework.
Our conclusions and discussion of the results appear in  Section~\ref{Conc}. 
Finally, some technical points are presented in the Appendix. 

\section{\label{gf}General framework}

In this section, we set up the necessary notation 
and review some of the most salient features of the 
PT-BFM framework, putting particular emphasis on the 
form of the SDE for the gluon propagator, 
and the various field-theoretic ingredients appearing in it. 

The (full) gluon propagator 
$\Delta^{ab}_{\mu\nu}(q)=\delta^{ab}\Delta_{\mu\nu}(q)$ in the renormalizable $R_\xi$ gauges is defined as
\be
i\Delta_{\mu\nu}(q)=- i\left[P_{\mu\nu}(q)\Delta(q^2)+\xi\frac{q_\mu q_\nu}{q^4}\right], \qquad
\Delta^{-1}_{\mu\nu}(q)=i\left[P_{\mu\nu}(q)\Delta^{-1}(q^2)+ (1/\xi)q_\mu q_\nu\right],
\label{prop}
\ee
with 
\be
P_{\mu\nu}(q)=g_{\mu\nu}- \frac{q_\mu q_\nu}{q^2}\,,
\ee
the dimensionless transverse projector, and $\xi$ the gauge fixing parameter.  The scalar cofactor $\Delta(q^2)$ appearing above is related to the all-order gluon self-energy $\Pi_{\mu\nu}(q)=P_{\mu\nu}(q)\Pi(q^2)$  through
\be
\Delta^{-1}({q^2})=q^2+i\Pi(q^2).
\label{defPi}
\ee
In addition, it is convenient to define the dimensionless function $J(q^2)$ as~\cite{Ball:1980ax}
\be
\Delta^{-1}({q^2})=q^2 J(q^2).
\label{defJ}
\ee
Evidently, $J(q^2)$ coincides with the {\it inverse} of the gluon dressing function, 
frequently considered in the literature.

\begin{figure}[!t]
\includegraphics[scale=.75]{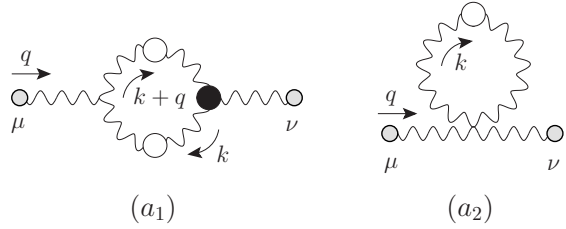}
\caption{\label{gSDE}The ``one-loop dressed'' gluon contribution to the  PT-BFM gluon self-energy. White (respectively, black) blobs represent connected (respectively, 1-PI) Green's functions; a gray circle on the external legs indicates background gluons. Notice that within the PT-BFM framework these 
two diagrams alone constitute a transverse subset of the full gluon SDE.}
\end{figure}

The starting point of our dynamical analysis is the SDE 
governing the gluon propagator.  
Within the PT-BFM framework that we employ~\cite{Aguilar:2006gr,Binosi:2007pi,Aguilar:2008xm,
Binosi:2008qk,Binosi:2009qm,Aguilar:2009ke,Cornwall:1981zr,Aguilar:2007fe,Cornwall:1989gv,Binosi:2002ft,Binosi:2003rr,Abbott:1980hw},  
one may safely truncate the SDE series down to its ``one-loop dressed version''  
containing gluonic contributions only, given by the diagrams $(a_1)$ and $(a_2)$ shown in Fig.~\ref{gSDE}.
Specifically, due to the special Feynman rules of the PT-BFM, 
and in particular the QED-like Ward identities satisfied by the 
fully-dressed vertices, gauge invariance remains exact, 
in the sense that 
the resulting (approximate) gluon self-energy $\Pi_{\mu\nu}(q)$
is still transverse, {\it i.e.}, 
\be
q^{\nu} \Pi_{\mu\nu}(q) = 0. 
\ee

\begin{figure}[!t]
\includegraphics[scale=.55]{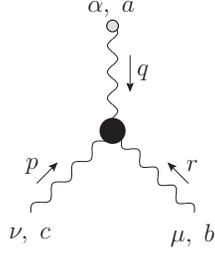}
\caption{\label{3g-vertex}The $BQQ$ three-gluon vertex.}
\end{figure}

The PT-BFM equation for the conventional propagator reads, in this case, 
\be
\Delta^{-1}(q^2){ P}_{\mu\nu}(q) = 
\frac{q^2 {P}_{\mu\nu}(q) + i\,[(a_1)+(a_2)]_{\mu\nu}}{[1+G(q^2)]^2},
\label{sde}
\ee
where
\bea
(a_1)_{\mu\nu}&=& \frac12\,g^2C_A
\int_k\!\gtreeb_{\mu\alpha\beta}(q,k,-k-q)\Delta^{\alpha\rho}(k)\Delta^{\beta\sigma}(k+q)\gfullb_{\nu\rho\sigma}(q,k,-k-q)\nonumber \\
(a_2)_{\mu\nu}&=&g^2C_A \left[g_{\mu\nu}\int_k\!\Delta^\rho_\rho(k)
+ \left(1/\xi-1\right)\int_k\!\Delta_{\mu\nu}(k)\right],
\label{gl-1ldr}
\eea
with $C_{{A}}$ being the Casimir eigenvalue of the adjoint representation
[$C_{{A}}=N$ for $SU(N)$], and the $d$-dimensional integral measure (in dimensional regularization) is defined according to
\be
\int_{k}\equiv\frac{\mu^{\epsilon}}{(2\pi)^{d}}\!\int\!\diff^d k.
\label{dqd}
\ee
The vertex $\gfullb$ is the fully dressed version of the trilinear vertex involving one background and two quantum gluons ($BQQ$ vertex for short, see Fig.~\ref{3g-vertex}); at tree-level (all momenta entering)
\be
\gtree_{\alpha\mu\nu}(q,r,p)= \gtreeb_{\alpha\mu\nu}(q,r,p)+ (1/\xi)
\Gamma^{{\rm P}}_{\alpha\mu\nu}(q,r,p),    
\ee
with
\bea
\gtree_{\alpha\mu\nu}(q,r,p)&=&g_{\mu\nu}(r-p)_\alpha  +
g_{\alpha\nu}(p-q)_\mu+g_{\alpha\mu}(q-r)_\nu,\nonumber \\
\gtreeb_{\alpha\mu\nu}(q,r,p)&=&g_{\mu\nu}(r-p)_\alpha  +
g_{\alpha\nu}(p-q+ r/\xi)_\mu+g_{\alpha\mu}(q-r- p/\xi)_\nu, \nonumber \\    
\Gamma^{{\rm P}}_{\alpha\mu\nu}(q,r,p)&=& g_{\alpha\mu} p_\nu - g_{\alpha\nu}r_\mu.
\label{deco}
\eea

\begin{figure}[!t]
\includegraphics[scale=.7]{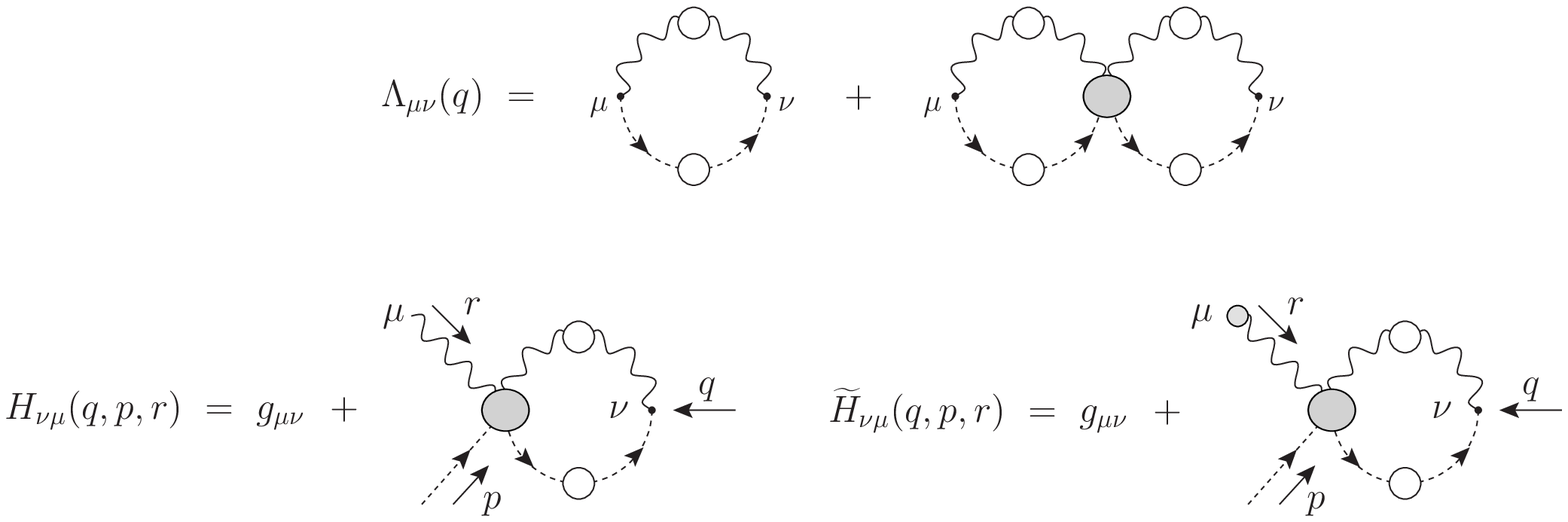}
\caption{\label{H-Htilde}Diagrammatic representation of the functions $\Lambda$, $H$ and, for later convenience, $\widetilde{H}$. Gray blobs represent 1-PI kernels with respect to vertical cuts.}
\end{figure}

Finally, the function $G(q^2)$ 
appearing in (\ref{sde}) is of central importance in this entire formalism. It is defined as 
the scalar co-factor of the $g_{\mu\nu}$ component of the special two-point function 
$\Lambda_{\mu\nu}(q)$, defined as (see also Fig.~\ref{H-Htilde})
\bea
\Lambda_{\mu\nu}(q)&=&-ig^2C_A\int_k\!\Delta_\mu^\sigma(k)D(q-k)H_{\nu\sigma}(-q,q-k,k)\nonumber\\
&=&g_{\mu\nu}G(q^2)+\frac{q_\mu q_\nu}{q^2}L(q^2),
\label{Lambda}
\eea
where we have introduced the ghost propagator $D^{ab}(q^2)=\delta^{ab}D(q^2)$, which 
is related to the ghost dressing function $F(q^2)$ through
\be
D(q^2)=  \frac{F(q^2)}{q^2}.
\ee

Notice that in the Landau gauge, an important exact (all-order) relation exists, 
linking $G(q^2)$ and  $L(q^2)$ to the 
ghost dressing function $F(q^2)$, namely~\cite{Grassi:2004yq,Aguilar:2009nf,Aguilar:2009pp,Aguilar:2010gm}  
\be
F^{-1}(q^2) = 1+G(q^2) + L(q^2).
\label{funrel}
\ee

In addition, the function $G(q^2)$ participates in a set of BRST-driven identities, 
known as Background-Quantum identities (BQIs)~\cite{Grassi:1999tp, Binosi:2002ez},  
obtained within the Batalin-Vilkovisky formalism~\cite{Batalin:1977pb,Batalin:1981jr}. These powerful identities 
relate  among each other the three types of gluon propagators that appear naturally in the BFM formalism, namely: 
({\it i}) the conventional gluon propagator (two quantum gluons entering, $QQ$), 
denoted by $\Delta(q^2)$; ({\it ii})
the background  gluon propagator (two background gluons entering, $BB$), 
denoted by $\widehat\Delta(q^2)$; and ({\it iii})
the mixed  background-quantum gluon propagator (one background and one quantum gluons entering, $BQ$), 
denoted by  $\widetilde\Delta(q^2)$. 
The corresponding BQIs are  
\bea
\Delta(q^2) &=& [1+G(q^2)]^2 \widehat\Delta(q^2),
\nonumber\\
\Delta(q^2) &=& [1+G(q^2)]\widetilde\Delta(q^2),
\nonumber\\
\widetilde\Delta(q^2) &=& [1+G(q^2)]\widehat\Delta(q^2).
\label{BQIs}
\eea
Notice that it is the first of these identities that allows the rewriting of the conventional SDE
into the PT-BFM form~(\ref{sde})~\cite{Binosi:2007pi,Aguilar:2008xm,Binosi:2008qk}.

For the rest of the article we will study the gluon SDE in the Landau gauge, $\xi=0$. 
The limit of Eq.~(\ref{sde}) as $\xi\to 0$ is rather subtle, and has been presented in~\cite{Aguilar:2008xm}. 
The final answer is 
\bea
{\widehat\Pi}^{\mu\nu}(q) &=& [(a_1)+(a_2)]_{\xi=0}^{\mu\nu}
\nonumber\\
&=&  g^2 C_A \sum_{i=1}^{5} A^{\mu\nu}_{i}(q) \,, 
\label{lan}
\eea
with     
\bea
A^{\mu\nu}_{1}(q)&=& \frac{1}{2}\int_k
\Gamma^{(0)\mu}_{\alpha\beta}P^{\alpha\rho}(k)P^{\beta\sigma}(k+q)\widetilde{\bqq}^\nu_{\rho\sigma} \Delta(k) \Delta(k+q),
\nonumber \\
A^{\mu\nu}_{2}(q)&=&  \int_k \! P^{\alpha\mu}(k) \frac{(k+q)^{\beta} \Gamma^{(0)\nu}_{\alpha\beta}}{(k+q)^2}\Delta(k),
\nonumber \\
A^{\mu\nu}_{3}(q)&=& \int_k \! P^{\alpha\mu}(k) \frac{(k+q)^{\beta} \widetilde{\bqq}^\nu_{\alpha\beta}}{(k+q)^2}\Delta(k),
\nonumber \\
A^{\mu\nu}_{4}(q)&=&-\frac{(d-1)^2}{d}g^{\mu\nu}\int_k \Delta(k), \nonumber \\
A^{\mu\nu}_{5}(q)&=& \int_k \frac{k^{\mu}(k+q)^{\nu}}{k^2 (k+q)^2}.
\label{thebees}
\eea
The vertex $\bqq$ appearing above is the fully-dressed PT-BFM vertex studied in detail in~\cite{Binosi:2011wi}, and which is related to the full $BQQ$ vertex $\widetilde{\Gamma}$ appearing in the BFM through
\be
\widetilde{\bqq}_{\alpha\mu\nu}(q,r,p)=\gfullb_{\alpha\mu\nu}(q,r,p)+(1/\xi)\Gamma^{\mathrm{P}}_{\alpha\mu\nu}(q,r,p).
\label{bqq}
\ee
Evidently, $\widetilde{\bqq}_{\alpha\mu\nu}(q,r,p)$ and $\gfullb_{\alpha\mu\nu}(q,r,p)$ differ only at tree level; 
specifically,  one sees immediately that 
\be
\widetilde{\bqq}_{\alpha\mu\nu}^{(0)}(q,r,p)  = \gtree_{\alpha\mu\nu}(q,r,p).
\ee
In the rest of this paper, we will refer indifferently to both $\gfullb$ and $\widetilde{\bqq}$ as the $BQQ$ vertex; in addition, in order to simplify the notation, we will drop the ``tilde'' superscript.

The vertex $\bqq$ satisfies a (ghost-free) WI when contracted with the momentum $q_\alpha$ of the 
background gluon, whereas it satisfies a STI when contracted with 
the momentum of the quantum gluons ($r_\mu$ or $p_\nu$). In particular, 
\bea
q^\alpha\bqq_{\alpha\mu\nu}(q,r,p)&=&p^2\bcj(p^2)P_{\mu\nu}(p)-r^2\bcj(r^2)P_{\mu\nu}(r),
\nonumber \\
r^\mu\bqq_{\alpha \mu \nu}(q,r,p)&=&F(r^2)\left[q^2\bcjb(q^2)P_\alpha^\mu(q)H_{\mu\nu}(q,r,p)-
p^2\bcj(p^2)P_\nu^\mu(p)\widetilde{H}_{\mu\alpha}(p,r,q)\right], \nonumber \\
p^\nu\bqq_{\alpha \mu \nu}(q,r,p)&=&F(p^2)\left[r^2\bcj(r^2)P_\mu^\nu(r)\widetilde{H}_{\nu\alpha}(r,p,q)-
q^2\bcjb(q^2) P_\alpha^\nu(q)H_{\nu\mu}(q,p,r)\right],
\label{STIs}
\eea
and the function $\bcjb$ is related to the conventional one defined in~(\ref{defJ}) precisely through the 
second equation in (\ref{BQIs}), namely
\be
\bcjb(q^2)=\left[1+G(q^2)\right]\bcj(q^2).
\ee
In addition, as shown in Fig.~\ref{H-Htilde}, the auxiliary ghost function $\widetilde{H}$ is the same as ${H}$ after converting  the external gluon leg into a background leg. An explicit form in terms of $J$, $\widetilde{J}$, $H$ and $\widetilde{H}$ of the (longitudinal) form factors characterizing this vertex has been obtained in~\cite{Binosi:2011wi} and reported in Appendix~\ref{app1}.

One may finally use Eq.~(\ref{defJ}) to re-express the relations~(\ref{STIs}) 
in terms of the (inverse) scalar functions $\Delta$, {\it i.e.}, 
\be
q^\alpha\bqq_{\alpha\mu\nu}(q,r,p) = \Delta^{-1}(p^2)P_{\mu\nu}(p) - \Delta^{-1}(r^2)P_{\mu\nu}(r), 
\label{WIal}
\ee
with analogous expressions holding for the remaining two STIs of (\ref{STIs}). 
At this level this appears as a simple rewriting, but 
this form of writing will facilitate the clarification of certain conceptual issues that become relevant  
when dynamical mass generation is turned on (see next section).

\section{\label{vert}Vertices in the presence of a dynamical mass}

In order to generate a dynamical mass without interfering with 
gauge invariance and the BRST symmetry, one must resort 
to the Schwinger mechanism~\cite{Schwinger:1962tn,Schwinger:1962tp}.
The general idea is to assume that a longitudinally coupled bound-state pole has been 
formed dynamically, which will modify the structure of the full vertices of the theory~\cite{Jackiw:1973tr,Cornwall:1973ts,Eichten:1974et,Jackiw:1973ha, Poggio:1974qs,Farhi:1982vt,Frampton:2008zz}.
This modification, in turn, will be responsible for obtaining massive type of solutions from the 
SDE of the gluon where this new vertices will be inserted~\cite{Aguilar:2008xm}. 
It is important to be very precise regarding the nature and role of the various ingredients that enter 
in the ensuing analysis. We will therefore devote this section to the development and elaboration  
of the various key concepts needed.

From the kinematic point of view 
we will describe the transition 
from a massless to a massive gluon propagator by carrying out the replacement  
(in Minkowski space)
\be
\Delta^{-1}(q^2) = q^2 J(q^2) \quad \longrightarrow\quad  \Delta_m^{-1}(q^2)=q^2 J_m(q^2)-m^2(q^2).
\label{massive}
\ee
The symbol $J_m$ indicates that effectively one has now a mass inside the corresponding expressions: for example, 
whereas perturbatively $J(q^2) \sim \ln q^2$,
after dynamical gluon mass generation has taken place, one has $J_m(q^2) \sim \ln(q^2+m^2)$.
As a consequence, since  $J_m$ will be the main component in the definition of  
the QCD effective charge~\cite{Aguilar:2009ke}, 
the presence of the mass term in the argument of its logarithm will tame the perturbative Landau pole
~\cite{Cornwall:1981zr,Aguilar:2009nf,Aguilar:2010gm}.   
Of course, as $q^2\to 0$,    $q^2 J_m(q^2) \to 0$; therefore, if we are to ensure that this procedure 
will give rise to a non vanishing IR value for the gluon propagator, {\it i.e.},~$\Delta_m^{-1}(0)\neq 0$, 
we must have that $m^2(0) \neq 0$.

\begin{figure}[!t]
\includegraphics[scale=.68]{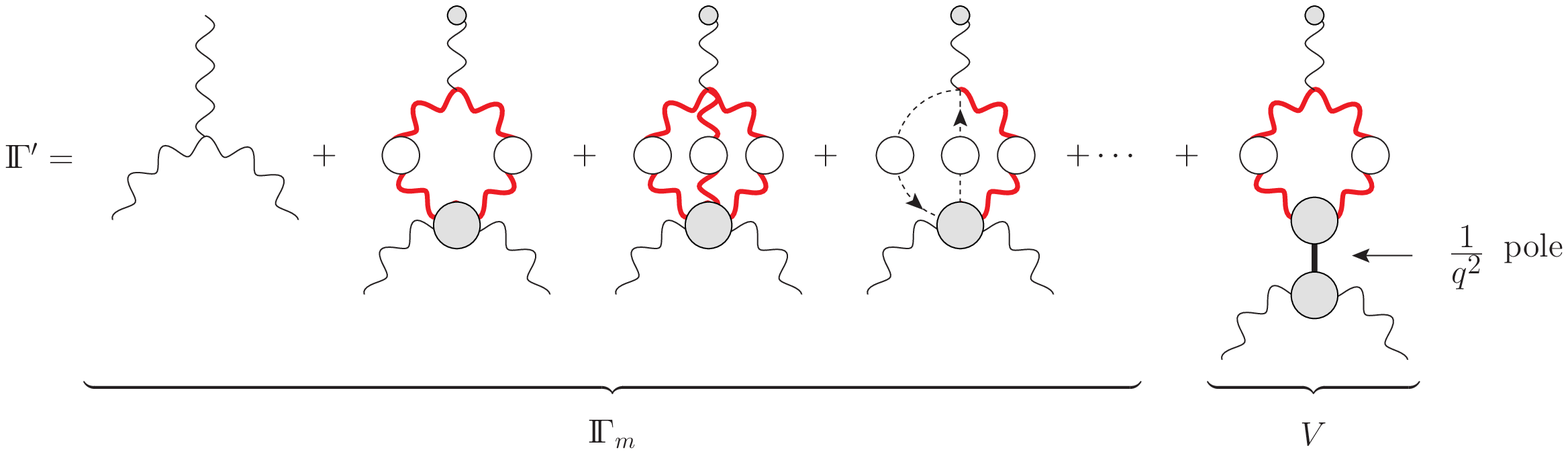}
\caption{\label{Gammaprime}The $\NV$ three-gluon vertex. 
Thick (red online) internal gluon lines indicates massive propagators $\Delta_m$, as explained in the text.}
\end{figure}

From the dynamical point of view,  it is clear 
that the full three-gluon vertex must be also 
appropriately modified~\cite{Jackiw:1973tr,Cornwall:1973ts,Eichten:1974et}. 
Specifically, we will consider a new vertex, to be denoted by $\NV$, and carry out the replacement 
\be
\bqq \quad \longrightarrow\quad   \NV = \bqq_m + \NP. 
\label{nv}
\ee
The new vertex $\bqq_m$ is given by the same (fully dressed) graphs 
that make up the SDE of the $BQQ$ vertex $\bqq$ (Fig.~\ref{Gammaprime});  
however, now all internal (virtual) fully-dressed gluon propagators   
are massive {\it i.e.},  in the non-pole part of the vertex SDE we have  
$\Delta \to \Delta_m$. In addition (or as a result thereof),  
$\bqq_m$ satisfies exactly the set of WI and STIs given in (\ref{STIs}), but with the 
replacement $J\to J_m$ throughout. 
So, the WI becomes 
\be 
q_\alpha\bqq_{m}^{\alpha\mu\nu}(q,r,p)=p^2\bcj_m (p^2)P^{\mu\nu}(p)-r^2\bcj_m(r^2)P^{\mu\nu}(r),
\label{wigm}
\ee
and exactly analogous expressions for the remaining STIs satisfied when $\bqq_{m}$ is contracted by either  $r$ or $p$. 
Note that all other Green's functions, such as 
$H$ and $\widetilde{H}$, must be replaced by the corresponding $H_m$ and $\widetilde{H}_m$, 
in the same sense as before (but we will suppress their `$m$' subindex throughout); thus, 
the diagrams defining these two ghost functions, shown in Fig.~\ref{H-Htilde}, will now contain massive internal gluon propagators. 

On the other hand,  the vertex $\NP$ represents the pole part of $\NV$; it is {\it totally longitudinally coupled}, 
{\it i.e.}, it vanishes identically when contracted by the three transverse projectors  
\be
P^{\alpha'\alpha}(q) P^{\mu'\mu}(r) P^{\nu'\nu}(p) \NP_{\alpha\mu\nu}(q,r,p)  = 0,
\label{totlon}
\ee 
and must satisfy the WI and STI of (\ref{STIs}), with the replacement $k^2 J(k) \to - m^2(k)$, {\it e.g.},
\be
q^\alpha \NP_{\alpha\mu\nu}(q,r,p)= - m^2(p^2)P_{\mu\nu}(p) + m^2(r^2)P_{\mu\nu}(r).
\label{winp}
\ee
Exactly analogous expressions will hold for the  
STIs satisfied when contracting with the  momenta $r$ or $p$. 

An explicit example of such a vertex (which, however, we will not use here),   
has been given in ~\cite{Cornwall:1985bg}, namely 
\bea
V^{\alpha\mu\nu}(q,r,p)&=&\frac{q^\alpha r^\mu (q-r)_\rho}{2q^2r^2}P^{\rho\nu}(p)m^2(p^2)-
\frac{p^\nu}{p^2}\left[m^2(r^2)-m^2(q^2)\right]P^\alpha_{\rho}(q)P^{\rho\mu}(r)\nonumber \\
&+&\frac{r^\mu p^\nu (r-p)_\rho}{2r^2p^2}P^{\rho\alpha}(q)m^2(q^2)-\frac{q^\alpha}{q^2}
\left[m^2(p^2)-m^2(r^2)\right]P^\mu_{\rho}(r)P^{\rho\nu}(p)\nonumber \\
&+&\frac{p^\nu q^\alpha (p-q)_\rho}{2q^2p^2}P^{\rho\mu}(r)m^2(r^2)-
\frac{r^\mu}{r^2}\left[m^2(q^2)-m^2(p^2)\right]P^\nu_{\rho}(p)P^{\rho\alpha}(q).
\eea
The totally longitudinal nature of this vertex is manifest\footnote{Note that 
this vertex is totally Bose symmetric, satisfying (\ref{winp}) with respect to all its legs; 
instead, the vertex considered here satisfies an STI with respect to the quantum legs ($r$ and $p$).}.

At this point it is clear that the full vertex $\NV$ will satisfy the same WI and STIs~(\ref{STIs}) satisfied by the $\bqq$ vertex before the introduction of any masses, but now with the replacement $\Delta \to \Delta_m$. 
Therefore, using Eqs.~(\ref{nv}), (\ref{wigm}), and (\ref{winp}), 
one gets for $\NV$ the WI
\bea
q_{\alpha}\NVindex(q,r,p) &=& q_{\alpha}\left[\bqq_m^{\alpha\mu\nu}(q,r,p) + \NP^{\alpha\mu\nu}(q,r,p)\right]
\nonumber\\
&=& [p^2\bcj_m (p^2) -m^2(p^2)]P^{\mu\nu}(p) - [r^2\bcj_m (r^2) -m^2(r^2)]P^{\mu\nu}(r)
\nonumber\\
&=& \Delta_m^{-1}(p^2) P^{\mu\nu}(p) - \Delta_m^{-1}(r^2) P^{\mu\nu}(r).
\label{winpfull}
\eea
Similarly
\bea
r_\mu\NVindex(q,r,p)&=&F(r^2)\left[\Delta_m^{-1}(q^2) P_\alpha^\mu(q)H_{\mu\nu}(q,r,p)-
\Delta_m^{-1}(p^2) P_\nu^\mu(p)\widetilde{H}_{\mu\alpha}(p,r,q)\right], \label{STI-2} \\
p_\nu\NVindex(q,r,p)&=&F(p^2)\left[ \Delta_m^{-1}(r^2) P_\mu^\nu(r)\widetilde{H}_{\nu\alpha}(r,p,q)-
\Delta_m^{-1}(q^2) P_\alpha^\nu(q)H_{\nu\mu}(q,p,r)\right].\hspace{.5cm}
\label{STI-1}
\eea

It is very important to emphasize that, 
even though the new (massive) WI is obtained from the old (massless) one through      
the replacement $\Delta \to \Delta_m$, the new vertex  $\NV$ is {\it not} obtained from 
the old one,  $\bqq$, by means of the same replacement {\it only}. Indeed, turning to the explicit expression for 
$\bqq$ given in the Appendix~\ref{app1}, it would certainly be wrong to 
use there the replacement $\Delta \to \Delta_m$ (or $J(q^2) \to \Delta_m(q^2)/q^2$). Instead, the correct procedure 
is that outlined above: the vertex $\bqq_m$ is indeed obtained from 
the expressions in the Appendix, by replacing $J \to J_m$ (but with no explicit mass terms); all explicit mass terms are next added through the totally longitudinally coupled non-perturbative vertex $\NP$. 

Actually, it is interesting to ponder about what would happen if one were to introduce the gluon 
mass through the (wrong) procedure of identifying the vertex $\NV$ by the simple replacement $\Delta \to \Delta_m$ carried out inside~$\bqq$. 
In such a case one would conclude (after some steps)  
that the self-consistency of the theory would force $m^{2}(q^2)$ to vanish identically. 
The precise way how this ``self-correction'' takes place is 
intimately related to the so-called ``seagull identity''~\cite{Aguilar:2009ke}, and will be discussed at the end of the 
Section~\ref{gml}. 

\section{\label{st}General features of the gluon mass equation}

\begin{figure}[!t]
\includegraphics[scale=.76]{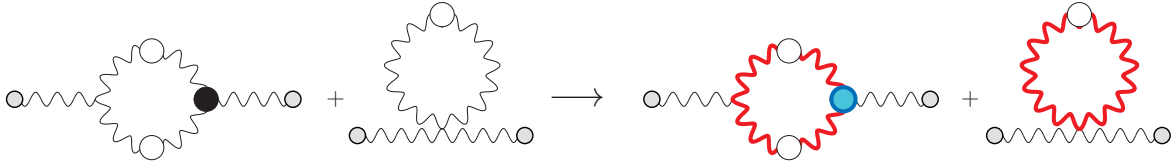}
\caption{\label{mod-gsde}Diagrammatic representation of the gluon one-loop dressed diagrams before and after dynamical gluon mass generation has taken place: the propagators and vertices on the rhs have now become massive.}
\end{figure}

Let us now consider the gluon SDE of Eq.~(\ref{sde}) under the light of the analysis presented in the previous section.  
After dynamical gluon mass generation has taken place, one needs to consider 
the modified SDE, which is obtained from~(\ref{sde}) after ({\it i}) replacing the $\Delta^{-1}$ 
appearing on the lhs with the $\Delta^{-1}_m$ of Eq.~(\ref{massive}), 
and ({\it ii}) replacing  $\Delta\to\Delta_m$ and $\bqq \to  \NV$ inside the integrals of the rhs
(see also Fig.~\ref{mod-gsde}).

From this new SDE one can obtain two separate equations, the first one governing the behavior of $J_m(q^2)$ 
[to be later involved in the definition of the effective charge, see Eq.~(\ref{RG-terms})] 
and the second one describing the dynamical mass $m^{2}(q^2)$.
The general idea is the following: the terms appearing on the rhs of the SDE may be separated systematically into two contributions, 
one that vanishes as $q\to 0$ and one that does not; the latter 
contribution must be set equal to the corresponding non-vanishing term on the lhs, namely $-m^{2}(q)$, 
while the former will be set equal to the vanishing term of the lhs, namely $q^{2} J_m(q^2)$, the so-called ``kinetic term''.   

Specifically (taking the trace of both sides of~(\ref{sde}) to eliminate the Lorentz indices), the rhs may be schematically cast 
in the form 
\be
q^{2} J_m(q^2) - m^{2}(q^2) =  
q^{2} \left[1+{\cal K}_1 (q^{2},m^2,\Delta_m)\right] + {\cal K}_2 (q^{2},m^2,\Delta_m),
\ee
such that $q^{2} {\cal K}_1 (q^{2},m^2,\Delta_m) \to 0$, as $q^{2}\to 0$, 
whereas ${\cal K}_2 (q^{2},m^2,\Delta_m)\neq 0$ in the same limit.   
Thus, for example, a term of the form  $q^{2} \int_k\! \Delta_m(k) \Delta_m(k+q)$ contributes to 
${\cal K}_1$, whereas a term of the form $m^{2}(q^{2}) \int_k\! \Delta_m(k) \Delta_m(k+q)$
should be assigned to ${\cal K}_2$. 
Then, the two equations determining  $J_m(q^2)$ and $m^{2}(q^2)$ will read (still Minkowski space) 
\bea
J_m(q^2) &=& 1+ {\cal K}_1 (q^{2},m^2,\Delta_m),
\nonumber\\
m^{2}(q^2) &=& - {\cal K}_2 (q^{2},m^2,\Delta_m).
\label{separ}
\eea

Of course, there is an obvious subtlety that must be addressed at this point. Specifically, 
one may easily envisage the possibility of a term that approaches zero as $(q^{2})^{a}$, with \mbox{$0<a<1$}. 
In this case, given that we must factor out a $q^{2}$ in order to obtain the equation for $J_m(q^2)$, 
such a term would furnish an IR divergent contribution to $J_m(q^2)$. This would be an undesirable feature, 
given that the $J_m(q^2)$ is intimately related to the effective charge of QCD, which   
is believed to be finite. The way to treat such a possibility is to state that, 
should such a term appear, it ought to be directly allotted (in its entirety, without factoring out 
a $q^{2}$) to the equation for $m^{2}(q^2)$. The presence of such a term in the mass equation 
will not affect the value of the mass at $q^2=0$, but will in general 
affect the shape of the resulting curve. 
Keeping this mathematical possibility in mind, 
let us point out that  
the terms emerging in the analysis of Section~\ref{gml}
have a very characteristic structure 
[see Eq.~(\ref{Ifirt}) and the related discussion], and, at least for them, 
the scenario contemplated above [ $(q^{2})^{a}$, with $0<a<1$] is not realized.

There is an additional point related to the 
mass equation, which is instrumental for the self-consistency of the 
entire approach.
Specifically, a crucial condition for the mechanism of dynamical 
gluon mass generation, developed in a series of articles~\cite{Cornwall:1981zr,Aguilar:2006gr,Binosi:2007pi,
Binosi:2008qk,Aguilar:2008xm}, 
is the cancellation of all seagull-type of divergences, {\it i.e.}, 
divergences produced by integrals of the type $\int_k \Delta(k^2)$, 
or variations thereof~\cite{Aguilar:2009ke}. 
The  precise cancellation of such terms proceeds by means of the 
identity~\cite{Aguilar:2009ke}
\be
\int_k\! k^2 \Delta'_m(k)+\frac{d}2\int_k\!\Delta_m(k)=0,
\label{seagull}
\ee 
where the ``prime'' denotes differentiation with respect to $k^2$, {\it i.e.}, 
$\Delta'_m(k) \equiv \partial{\Delta_m}(k^2)/\partial k^2$.
Thus, all the ingredients entering into the SDE (most importantly, the vertex) 
must be such that, 
after taking the limit of the 
SDE as $q^2\to 0$, all seagull-type contributions must conspire to 
appear exactly in the combination given on the lhs of Eq.~(\ref{seagull}). 

The relevance and function of the identity (\ref{seagull}) becomes evident when  
we consider the term $I(q)$, given by 
\be
-iI(q) = \int_k\! k^2 \Delta_m (k)\Delta_m (k+q)  \frac{(k+q)^2J_m(k+q)-k^2J_m(k)}{(k+q)^2-k^2} 
+ c \int_k \Delta_m(k),
\ee
with $c$ (for the moment) an arbitrary real number. This term  appears  naturally
in the PT-BFM framework, and in fact we will find it in the case of the Landau gauge studied in the next section.

Using Eq.~(\ref{massive}) one may then rewrite $I(q)$ as 
\be
I(q) = I_1(q) + I_2(q),
\ee
with 
\bea
-iI_1(q) &=& - \int_k\! k^2 \frac{\Delta_m (k+q)- \Delta_m (k)}{(k+q)^2-k^2} + c \int_k \Delta_m(k)
\nonumber\\
&=& - \left[\int_k\! k^2 \frac{\Delta_m (k+q)- \Delta_m (k)}{(k+q)^2-k^2} + \frac{d}2\int_k\!\Delta_m(k)\right]
+ \left(c+\frac{d}2\right) \int_k \Delta_m(k),
\eea
and 
\be
-iI_2(q) = \int_k\! k^2 \Delta_m (k)\Delta_m (k+q) \frac{m^2 (k+q)- m^2 (k)}{(k+q)^2-k^2}.
\label{I2}
\ee
In order to establish how the above terms must be assigned among the ${\cal K}_1$ and ${\cal K}_2$ introduced above, 
let us now take their limit as $q^2\to 0$. Carrying out the appropriate Taylor expansions 
[see Eq.~(\ref{tayl})], one finds
\bea
-iI_1(0) &=& - \left[ \int_k\! k^2 \Delta'_m(k) +\frac{d}2\int_k\!\Delta_m(k) \right]
+ \left(c+\frac{d}2\right) \int_k \Delta_m(k) 
\nonumber\\
&=& \left(c+\frac{d}2\right) \int_k \Delta_m(k),
\eea
where in the second step we have employed Eq.~(\ref{seagull}), and 
\be
-iI_2(0) = \int_k\! k^2 \Delta^2_m (k) [m^2(k)]'.
\ee
Thus, according to the rules introduced above, the contribution of $I(q)$ to the kinetic term is 
\be
iI_{\rm kt}(q) = \int_k\! k^2 \frac{\Delta_m (k+q)- \Delta_m (k)}{(k+q)^2-k^2} + \frac{d}2\int_k\!\Delta_m(k),
\label{Ikt}
\ee
given that $I_{\rm kt}(0)=0$, while the contribution of $I(q)$ to the mass equation is 
\be
-iI_{m^2}(q) = \int_k\! k^2 \Delta_m (k)\Delta_m (k+q) \frac{m^2 (k+q)- m^2 (k)}{(k+q)^2-k^2} + \left(c+\frac{d}2\right) 
\int_k\Delta_m(k).
\label{mcon}
\ee
It is clear now that the second term on the rhs  of (\ref{mcon}) is quadratically divergent (and of the seagull type).  The only way to 
avoid this divergence is if the coefficient multiplying $\int_k\Delta_m(k)$ vanishes, {\it i.e.}, if $c=-d/2$.
It turns out that, by virtue of the PT-BFM Feynman rules, and the fact that gauge invariance is preserved at every level of this approximation, the coefficient $c$ comes out precisely equal to $-d/2$; we emphasize that this result can be realized {\it only} within the PT-BFM framework. 
Thus, after the seagull cancellation, one is left with the first term only, which is perfectly convergent, 
provided that the mass decreases in the deep ultraviolet. As we will see in the next section, 
in the Landau gauge this term accounts for the bulk of the gluon mass equation. 

Even though the term $I_{\rm kt}(q)$ of Eq.~(\ref{Ikt}) does not appear in the rest of our analysis, 
it is important to gain some further intuition on its structure and its 
behavior for small values of $q^2$, especially in the light of the discussion following Eq.~(\ref{separ}).

To this end, let us introduce spherical coordinates through the 
definitions \mbox{$q^2 =x$}, \mbox{$k^2 =y$}, \mbox{$(k+q)^2 =z$}; 
we then have that \mbox{$z= y+x+2 \sqrt{xy} \cos\theta $}, and we define \mbox{$ w \equiv (k+q)^2-k^2 = z-y= x+2 \sqrt{xy} \cos\theta$}. 
The $d$-dimensional integral measure will  read in this case
\be
\int_k\ =\ \frac1{(2\pi)^d}\frac{\pi^{\frac{d-1}2}}{\Gamma\left(\frac{d-1}2\right)}\int_0^\pi\!\diff\theta\sin^{d-2}\theta\int_0^\infty\!\diff y\,y^{\frac d2-1},
\label{d-measure}
\ee
and we finally recall the elementary integral 
\be
\int_{0}^{\pi}\!\diff\theta\,\sin^m\theta \cos^n\theta =
\left\{
\begin{array}{ll}
\frac{\Gamma\left(\frac{m+1}{2}\right) 
\Gamma\left( \frac{n+1}{2}\right) }{\Gamma\left( \frac{m+n+2}{2}\right)},\quad & n\ \mathrm{even} \\
0 , & n\ \mathrm{odd}
\end{array}
\right.
\label{intsin}
\ee

It turns out that the $I_{\rm kt}(q)$ of Eq.~(\ref{Ikt}) 
may be expanded systematically as a power series in $q^2$. 
To see this in detail, we consider the Taylor expansion of an arbitrary finite function $f(z)$ around $w=0$,  given by    
\be 
\frac{f(z)- f(y)}{w} = f^{\prime}(y) + 
\frac{w}{2!}   f^{\prime\prime}(y) + \frac{w^2}{3!}  f^{\prime\prime\prime}(y) + ...
\label{tayl}
\ee
where the primes denote differentiations with respect to $y$ (evidently we are assuming finite derivatives in the origin). 
Then, under the integral sign on the rhs of Eq.~(\ref{Ikt})  
one must collect pieces of a given order in $q^2$ from the various powers of $w$, 
using (\ref{intsin}).  

It is clear that 
when the term  $f^{\prime}(y)$ 
on the rhs of (\ref{tayl}) is inserted into the integral,    
it generates the seagull identity~(\ref{seagull}); all the remaining terms will be proportional to positive powers of $w$, and 
thus, $I_{\rm kt}(0)=0$. For example, the $q^2$ term in this expansion is obtained by appropriately 
combining contributions proportional to $f^{\prime\prime}(y)$ and $f^{\prime\prime\prime}(y)$. 
Using again~(\ref{intsin}), after a sequence of partial integrations, we find 
\be
iI_{\rm kt}(q) =\frac{q^2}{6} \left(\frac{d-4}{d} \right) \int_k k^2 \Delta''_m(k) + {\cal O}(q^4).
\label{Ifirt}
\ee


In order to check the validity of Eq.~(\ref{Ifirt})
let us compute 
$I_{\rm kt}(q)$  for the simple case of a massive propagator with a ``hard'' (momentum-independent) mass
\be
\Delta_m(q) = \frac{1}{q^2 -m^2}.
\label{hardm}
\ee
The integrand in the first integral on the rhs of Eq.~(\ref{mcon}) simplifies to 
\be
k^2 \, \frac{{\Delta}_m(k+q) - {\Delta}_m(k)}{(k+q)^2-k^2}= 
- \frac{k^2}{(k^2 -m^2 )[(k+q)^2 - m^2]}.
\label{ff1}
\ee
Then, using the dimensional regularization identity 
\be
2 m^2\int_k \frac{1}{(k^2-m^2)^2} = (d-2)\int_k \frac{1}{k^2-m^2},
\label{id2}
\ee
it is relatively straightforward to demonstrate that 
\be
I_{\rm kt}(q) = \frac{m^2}{16\pi^2}   
\int_{0}^{1}\!\diff x \ln\left(1 + \frac{q^2 x(x-1)}{m^2} \right).
\ee
Evidently, $I_{\rm kt}(0) =0$, as expected, while the expansion of the logarithm furnishes immediately the 
result 
\be
I_{\rm kt}(q) = -\frac{1}{16\pi^2} \frac{q^2}{6} \, +  {\cal O}(q^4).
\label{Iex}
\ee

On the other hand, substitution into the general formula (\ref{Ifirt})
of the propagator in (\ref{hardm}) yields 
\be
I_{\rm kt}(q) =-i \frac{q^2}{6} \left(\frac{d-4}{d} \right) 2 \int_k \frac{k^2}{(k^2-m^2)^3}  + {\cal O}(q^4).
\label{Ifirt1}
\ee
In dimensional regularization, around $d=4$, we have that $d=4-\epsilon$, and therefore, only the 
divergent part of the integral contributes, {\it i.e.}, 
\bea
I_{\rm kt}(q) &=& -i \frac{q^2}{6} \left(\frac{-\epsilon}{d} \right) 2 
\left[\frac{1}{16\pi^2} \left(\frac{id}{4}\right) \left(\frac{2}{\epsilon}\right)\right]   + {\cal O}(q^4)
\nonumber\\
&=& -\frac{1}{16\pi^2} \frac{q^2}{6} \,+ {\cal O}(q^4),
\label{Ifirt2}
\eea
which indeed coincides with (\ref{Iex}).

Notice finally that the main contribution to the kinetic term does not originate from $I_{\rm kt}(q)$,  
but rather from a term of the form 
\be
q^2 \int_k\! \frac{\Delta_m(k+q)- \Delta_m(k)}{(k+q)^2-k^2},
\ee
which, for the simple massive propagators of (\ref{hardm}) may be easily calculated, 
giving rise to the standard logarithmic correction associated 
with the RG, with the additional feature of  being IR safe due to the presence of the mass in the argument of the logarithm.

\section{\label{gml}The gluon mass equation in the Landau gauge}

We now proceed to the actual derivation of the explicit form of the mass equation in the Landau gauge. 
Specifically, in this gauge 
the rhs of  Eq.~(\ref{sde}) will be given by the terms $A_i$ listed in  Eq.~(\ref{thebees}), 
where now we must carry out the replacements  $\Delta \to \Delta_m$ and $\bqq \to \NV$.

Following the rules explained in the previous section, 
and defining 
\be
A_i(q)=\mathrm{Tr}\,[A_i^{\mu\nu}(q)], 
\ee
the mass equation is given by 
\be
m^2(q^2)=-i\frac{g^2C_A}{d-1}\frac{\left[\sum_{i=1}^5 A_i(q)\right]_{m^2}}{[1+G(q^2)]^2},
\label{me-gen}
\ee
and therefore, one should determine the closed form of the quantities $[A_i(q)]_{m^2}$.

There is a simple observation, particular to this gauge,
which simplifies the entire procedure considerably.
Specifically, in the Landau gauge, the derivation of the 
gluon mass equation does not require the knowledge of 
the closed form of the vertex $\NP$, which captures the effects of the  
massless bound-state poles. 

To see why this is so, let us first note that the vertex $\NP$ appears only in the terms 
$A^{\mu\nu}_{1}(q)$ and $A^{\mu\nu}_{3}(q)$, the only place where the replacement $\bqq \to \NV$ 
may be carried out. 
Given that the vertex $\NV_{\nu\alpha\beta}$ appearing 
in the term  $A^{\mu\nu}_{3}(q)$ 
is contracted by $(k+q)^{\beta}$, the result of this operation is the STI  satisfied by  $\NV$, 
namely 
\be
p^\nu\NV_{\alpha \mu \nu}=F(p^2)\left[\Delta_m^{-1}(r^2)P_\mu^\nu(r)\widetilde{H}_{\nu\alpha}(r,p,q)-
\widetilde{\Delta}_m^{-1}(q^2) P_\alpha^\nu(q)H_{\nu\mu}(q,p,r)\right]. 
\label{STI-2full}
\ee
whose validity assumes the existence of $\NP$ but does not depend on the details of its closed form.

As for the term $A^{\mu\nu}_{1}(q)$ 
one starts by noticing that ({\it i}) the $\NP$ is already contracted by two projection operators 
$P^{\alpha\rho}(k)P^{\beta\sigma}(k+q)$ and ({\it ii}) 
since in the PT-BFM formulation the truncated ${\widehat\Pi}^{\mu\nu}(q)$ (defined in terms 
of $A_1 - A_5$)  is transverse, one may contract both sides of Eq.~(\ref{lan}) by the projection operator 
$P^\nu_{\nu'}(q)$ {\it for free}, {\it i.e.}, write 
\bea
{\widehat\Pi}^{\mu\nu}(q) &=& {\widehat\Pi}^{\mu\nu'}(q) P^\nu_{\nu'}(q) 
\nonumber\\
&=& g^2 C_A \sum_{i=1}^{5} A^{\mu\nu'}_{i}(q) P^\nu_{\nu'}(q).
\label{lanpr}
\eea
The main effect of this operation, as far 
as the term $A^{\mu\nu}_{1}(q)$ is concerned, is to trigger Eq.~(\ref{totlon}), 
and so, all explicit reference to $\NP$ vanishes.

In order to forestall any possible confusion, we hasten to emphasize that one should not conclude from the above argument that the existence of the vertex  $\NP$ is irrelevant for the entire construction. 
On the contrary, the vertex  $\NP$ is crucial for the implementation of this particular approach. 
In particular, if the $\NP$ did not exist ({\it i.e.}, if it were vanishing identically) 
the WI of (\ref{winp}) would be invalidated, and, as a result, ${\widehat\Pi}^{\mu\nu}(q)$ 
would fail to be transverse, in which case, evidently, one could no longer  
contract both sides of Eq.~(\ref{lan}) by the projection operator 
$P^\nu_{\nu'}(q)$ for free. 

We can now proceed with the actual calculation. 
It is clear that the term  $A^{\mu\nu}_{5}(q)$ cannot possibly contribute to the mass equation, since 
$A^{\mu\nu}_{5}(0)=0$. Furthermore, with the exception of $A^{\mu\nu}_{3}(q)$, which will yield a direct contribution, the remaining three terms $A^{\mu\nu}_{1}(q)$, $A^{\mu\nu}_{2}(q)$, and $A^{\mu\nu}_{4}(q)$  
contribute to the mass equations an amount that arises as the deviation from the 
seagull cancellation, {\it i.e.}, they furnish a term analogous to the $I_2$ of Eq.~(\ref{I2}).  

To see this, let us first retain the contributions of these three terms that 
survive {\it individually} the  $q^2\to 0$ limit. The term  $A_1$ reads 
\be
A_1(q)=\frac{1}{2}\int_k
\Gamma^{(0)}_{\mu\alpha\beta} P^{\alpha}_{\rho}(k) P^{\beta}_{\sigma}(k+q)\bqq_{m}^{\nu'\rho\sigma}P_{\nu'}^{\mu}(q)\Delta_m(k)\Delta_m(k+q).
\label{A1}
\ee
Now, using for the vertex $\bqq_{m}$ the tensor decomposition~(\ref{Ls}) with $(\alpha,\mu,\nu)\to(\nu',\rho,\sigma)$ and $r\to k$, $p\to-k-q$, it is straightforward to establish that the tensorial structures $\ell_2$, $\ell_5$ and $\ell_8$ will be annihilated 
by the transverse projectors appearing in~(\ref{A1}), while, ignoring terms that will again vanish due to the transverse projectors, $\ell_1$, $\ell_3$, $\ell_7$, and $\ell_9$  are at least of order $q$. Finally, since $\ell_{10}=0$, we find the result
\bea
\bqq_{m}^{\nu'\rho\sigma}(q,k,-k-q)&=&2k^{\nu'}g^{\rho\sigma}\left[X_4(q,k,-k-q)+k^2X_6(q,k,-k-q)\right]+{\cal O}(q)\nonumber \\
&=& 2k^{\nu'}g^{\rho\sigma}\frac{(k+q)^2J_m(k+q)-k^2J_m(k)}{(k+q)^2-k^2}+{\cal O}(q).
\eea
In addition, since 
\be
\Gamma^{(0)}_{\mu\alpha\beta}(q,k,-k-q)k^{\nu'}P_{\nu'}^{\mu}(q)=2g_{\alpha\beta}\left[k^2-\frac{(k\cdot q)^2}{q^2}\right]+{\cal O}(q),
\ee
we finally obtain 
\be
A_1(q)=2(d-1)\int_k\!\left[k^2-\frac{(k\cdot q)^2}{q^2}\right]\frac{(k+q)^2J_m(k+q)-k^2J_m(k)}{(k+q)^2-k^2}\Delta_m(k)\Delta_m(k+q)+{\cal O}(q).
\label{one}
\ee

Similarly, from $A^{\mu\nu}_{2}(q)$ we obtain
\be
A_2(q)= -\int_k \! \left[d-2 + \frac{(k \cdot q)^2}{k^2 q^2}\right ] \frac{k^2\, \Delta_m(k) }{(k+q)^2},
\label{two}
\ee
while $A^{\mu\nu}_{4}(q)$ contributes simply 
\be
A_4(q) = -\frac{(d-1)^3}{d} \int_k \Delta_m(k).
\label{four}
\ee

The terms in Eqs.~(\ref{one}), (\ref{two}) and~(\ref{four}) are individually non-vanishing as $q^2\to 0$, 
but their final contribution to the mass equation is controlled by the seagull identity, which forces a large part of their sum to vanish, thus 
reassigning them to the kinetic term. 
Specifically, if we use Eq.~(\ref{massive}) to substitute the terms containing $J_m$ in the numerator of the integral on the rhs of Eq.~(\ref{one}),
[{\it i.e.}, $k^2 J(k) = \Delta_m^{-1}(k) + m^2(k)$]
the sum of these three terms gives 
\be
[A_1 + A_2 + A_4](q) =  [A_1 + A_2 + A_4]_{\rm kt}(q) + [A_1 + A_2 + A_4]_{m^2}(q),
\ee
where
\bea
[A_1 + A_2 + A_4]_{\rm kt}(q) &=& - 2(d-1) \int_k\!\left[k^2-\frac{(k\cdot q)^2}{q^2}\right]
\frac{\Delta_m (k+q)- \Delta_m (k)}{(k+q)^2-k^2}
\nonumber\\
&-& \int_k \! \left[d-2 + \frac{(k \cdot q)^2}{k^2 q^2}\right ] \frac{k^2\, \Delta_m(k) }{(k+q)^2}
-\frac{(d-1)^3}{d} \int_k \Delta_m(k),
\label{A1kt}
\eea
and leaves as residual contribution
\be
[A_1 + A_2 + A_4)]_{m^2}(q)
=2(d-1)\int_k\!\left[k^2-\frac{(k\cdot q)^2}{q^2}\right]\frac{m^2(k+q)-m^2(k)}{(k+q)^2-k^2}\Delta_m(k)\Delta_m(k+q).
\label{A1m}
\ee

It is now easy to verify that, by virtue of the seagull identity, the rhs of~(\ref{A1kt}) vanishes as  $q^2\to 0$.
Indeed
\be
[A_1 + A_2 + A_4]_{\rm kt}(0) =  - 2(d-1) \int_k\! \sin^2 {\theta}\, k^2 \Delta'_m(k) -  
\int_k\! \left [(d-1) - \sin^2 {\theta} + \frac{(d-1)^3}{d}\right] \Delta_m(k),
\label{A124seag}
\ee
and using that [see also Eqs.~(\ref{d-measure}) and ~(\ref{intsin}) above]
\be
\int_k\! \sin^2 {\theta} \, f(k) = \frac{d-1}{d} \int_k\! f(k), 
\ee
it is elementary to demonstrate that the rhs is exactly proportional to the 
expression on the rhs of Eq.~(\ref{seagull}), and therefore vanishes. 

We next consider the term $A_{3}$.
After taking the trace we find
\be
A_{3}(q)= \int_k \! P^{\alpha\mu}(k) \frac{(k+q)^{\beta} \NV_{\nu'\alpha\beta}}{(k+q)^2}P^{\nu'}_\mu(q)\Delta_m(k).
\label{A3}
\ee
When inserted into the expression for $A_{3}(q)$, 
the first term on the rhs of~(\ref{STI-1}) will give the result
\be
\int_k\frac{F(k+q)}{(k+q)^2}\widetilde{a}(q,-k-q,k)+{\cal O}(q),
\ee
which contributes to the kinetic term, since in the $q\to0$ limit vanishes due to the second identity in~(\ref{constr}), which in this limit gives~\cite{Binosi:2011wi}
\be
\widetilde{a}(0,-k,k)=F^{-1}(k).
\ee
The second term on the rhs of~(\ref{STI-1}) yields instead a surprisingly simple contribution to the mass equation. Specifically, using the definition~(\ref{Lambda}) we obtain
\bea
[A_{3}]_{m^2}(q)  &=&
\widetilde{m}^2(q^2)\int_k\!\frac{F(k+q)}{(k+q)^2}\Delta_{\mu}^\rho(k)H_{\sigma\rho}(q,-k-q,k)P^{\sigma\mu}(q)
\nonumber \\
&=&\widetilde{m}^2(q^2)\frac{i\Lambda_{\sigma\mu}(q)}{g^2C_A}P^{\sigma\mu}(q)
\nonumber \\
&=&i\frac{d-1}{g^2C_A}\,\widetilde{m}^2(q^2)G(q^2).
\label{A3mtilde}
\eea
On the other hand, the second of the background quantum identities~(\ref{BQIs}) implies (see also Appendix~\ref{app2} for an alternative derivation of this result)
\be
\widetilde{m}^2(q^2)=[1+G(q^2)]m^2(q^2),
\label{mtilde-m-1}
\ee
so that one finally finds the contribution
\be
[A_{3}]_{m^2}(q) =i\frac{d-1}{g^2C_A}G(q^2)[1+G(q^2)]m^2 (q^2).
\label{A3m}
\ee

The next step is to substitute the above results on the rhs of the mass equation  of Eq.~(\ref{me-gen}). 
In doing so, we move to the Euclidean space, by setting $\int_k\!=\mathrm{i}\!\int_{k_\mathrm{\s E}}$ and $q^2_\mathrm{\s E} = -q^2$, 
and using  
\be
\Delta_\mathrm{\s E}(q^2_\mathrm{\s E})=-\Delta(-q^2_\mathrm{\s E}); \qquad m^2_\mathrm{\s E}(q^2_\mathrm{\s E})=  m^2(-q^2_\mathrm{\s E});
\qquad G_\mathrm{\s E}(q^2_\mathrm{\s E})= G(-q^2_\mathrm{\s E}). 
\ee
Then, from Eq.~(\ref{A1m}) and~(\ref{A3m}), 
we arrive at the final form of the mass equation, namely
\be
m^2(q^2)=\frac{2g^2C_A}{1+G(q^2)}\int_k\!\left[k^2-\frac{(k\cdot q)^2}{q^2}\right]
\frac{m^2(k+q)-m^2(k)}{(k+q)^2-k^2}\Delta_m(k)\Delta_m(k+q)\,,
\label{me-final}
\ee
where we have suppressed the suffix ``E''. 

Finally, we are now in position to address  
the question posed at the end of Section~\ref{vert}, namely what would happen if we were 
to introduce the gluon mass by the simple replacement $q^2 J(q^2) \to \Delta_m^{-1}(q^2)$ carried out inside~$\bqq$, {\it i.e.}, 
without resorting explicitly to  the vertex $\NP$ (with the crucial properties assigned to it).
The basic observation is that the main bulk of the mass equation, namely the rhs of Eq.~(\ref{A1m}), 
emerges as a residual contribution that survives the seagull cancellation. 
However, within this hypothetical scenario, the term $A_1(q)$ in Eq.~(\ref{one}) would be instead given by 
\bea
A_1(q) &=& 2(d-1)\int_k\!\left[k^2-\frac{(k\cdot q)^2}{q^2}\right]
\frac{\Delta_m^{-1}(k+q)- \Delta_m^{-1}(k)}{(k+q)^2-k^2}\Delta_m(k)\Delta_m(k+q)
\nonumber\\
&=& - 2(d-1)\int_k\!\left[k^2-\frac{(k\cdot q)^2}{q^2}\right]\frac{\Delta_m(k+q)-\Delta_m (k)}{(k+q)^2-k^2}\,, 
\label{onehyp}
\eea
thus, participating in the cancelation of Eq.~(\ref{A124seag}), as before, but leaving no residual contribution, {\it i.e.}, 
$[A_1 + A_2 + A_4]_{m^2}(q) =0$. 
Then, the only contribution to the rhs of the mass equation would be that of $[A_{3}]_{m^2}(q)$ in Eq.~(\ref{A3m}); this contribution 
would be still there, because within this alternative scenario the full vertex $\NV$
is still assumed to satisfy the full STIs  of Eq.~(\ref{STI-1}) [but with no reference to $V$].  
Therefore, the resulting mass equation [the equivalent of Eq.~(\ref{me-final})]
would read 
\be
\frac{m^2(q^2)}{\left[1+G(q^2)\right]}=0,
\label{mhyp}
\ee
which would simply imply $m^2(q^2)=0$, {\it i.e.}, no dynamical mass generation.

\section{\label{numan}Numerical analysis}

In this section, we will first derive an approximate version of the mass equation~(\ref{me-final}), 
which will facilitate the numerical treatment while retaining 
the main features of the full equation.   
Then, using as input for  the functions $\Delta(q^2)$ and  $F(q^2)$ [appearing in (\ref{me-final})]  
the available lattice data, 
we solve the equation numerically for the  gauge groups $SU(2)$ and $SU(3)$, thus obtaining 
the (approximate) form of $m^2(q^2)$. Then, using Eq.~(\ref{massive}), together with 
the $\Delta(q^2)$ 
of the lattice and the $m^2(q^2)$ obtained from the mass equation, 
we will extract the  
(approximate) form of $J_m(q^2)$. As a basic application, 
these ingredients will be subsequently combined to form 
the gluon mass entering in 
the RG-invariant combination  
associated with the definition of a non-Abelian effective charge.

\subsection{Approximate version of the mass equation}

We now proceed to the analysis of the mass equation~(\ref{me-final}). The difficulty in dealing with this equation in its full version resides in the fact that  the unknown function $m^2$ appearing on the rhs depends on both  the angular and the radial coordinates ($\theta$ and $y$, respectively).  To circumvent this problem we will employ certain standard approximations, in order to eliminate the angular integration. However, 
before embarking into the derivation  of the approximate version of~(\ref{me-final}), 
we can extract useful of information about the global behavior of $m^2$ from its $q^2\to0$ limit.

Specifically, let us employ the notation introduced in~(\ref{d-measure}), 
and consider the limit  of Eq.~(\ref{me-final}) as  $q^2\to0$. 
Since it is known that $L(0) = 0$ in four dimensions~\cite{Aguilar:2009nf}, Eq.~(\ref{funrel}) implies that $1+G(0) = F^{-1}(0)$, so that we get  
\bea
m^2(0)&=&\frac{3}{2} g^2 C_A F(0)\int_k\!k^2[m^2(k)]'\Delta^2(k)\nonumber \\
&=&-3g^2C_A F(0) \int_k\!m^2(k)\Delta(k)\left[k^2\Delta(k)\right]'.
\label{me-qtozero}
\eea 
Obviously, in the kernel of above equation there is no dependence on $\theta$,  
so that the angular integral can be done exactly, and one is left with  the final equation  
\bea
m^2(0) 
&=& -\frac{3C_A}{8\pi}\alpha_s F(0) \int_0^\infty\!\diff y\,m^2(y) [y^2\Delta^2(y)]',
\label{m20}
\eea 
where \mbox{$\alpha_s=g^2/4\pi$} and, as usual, \mbox{$y=k^2$} (the prime indicates now derivatives with respect to $y$).  

Equations~(\ref{me-qtozero}) and~(\ref{m20}) furnish a rather 
interesting constraint on the structure of the full gluon propagator. Indeed, it is clear that due to 
the positive sign in front of the first line of Eq.~(\ref{me-qtozero}), solutions of~(\ref{me-final}) 
leading to a positive $m^2(0)$ cannot be monotonically decreasing; or, seeing it from the point of view of Eq.~(\ref{m20}), 
the kernel $[y^2\Delta^2(y)]'$ must reverse sign and display a ``sufficiently deep'' negative region 
at intermediate momenta, in order to obtain $m^2(0)>0$. This is a highly non-trivial requirement, 
because, to the best of our knowledge, there is no {\it a priori} fundamental reason 
why the full gluon propagator propagator should show this particular behavior. 

We now proceed to the derivation of an approximate version of~(\ref{me-final}) 
that will reproduce in the $q^2\to0$ limit Eq.~(\ref{me-qtozero}),  
and therefore implement the important constraint that this equation entails. 

Let us then
denote by $R(q)$ the integral appearing on the rhs of (\ref{me-final}); using the simple identity  
\be
(k\cdot q)^2 = \frac{1}{4} \left\{[(k+q)^2-k^2]^2 - 2  q^2 [(k+q)^2-k^2] + (q^2)^2 \right\},
\ee 
we see that the second term above, when inserted back into $R(q)$,  vanishes upon integration,  and therefore one is left with 
\be
R(q) = R_1(q) + R_2(q),
\ee
where
\bea
R_1(q) &=& \int_k\!\left(k^2-\frac{q^2}{4}\right) \frac{m^2(k+q)-m^2(k)}{(k+q)^2-k^2}\Delta_m(k) \Delta_m(k+q),
\nonumber\\
R_2(q) &=& - \frac{1}{2q^2} \int_k\! m^2(k) [(k+q)^2-k^2]\Delta_m(k) \Delta_m(k+q).
\eea 

To cast  $R_1(q)$ and $R_2(q)$ into a form suitable for solving the corresponding dynamical equation, we first introduce the by now familiar spherical coordinates and then
split the radial integration into two intervals
\be
\int_0^{\infty}\! \diff y=  \int_0^{x}\! \diff y+ \int_x^{\infty}\! \diff y,
\ee
so that in the second integral since $y>x$ always, we can expand the integrand according to~(\ref{tayl}). 
Proceeding in this way, and observing that partial integration gives 
\be
\int_x^{\infty}\!\diff y\, y^2 [m^2(y)]'\Delta^2(y) = - m^2(x) x^2  \Delta^2(x) -  \int_x^{\infty}\!\diff y\,  m^2(y) [y^2\Delta^2(y)]'  
\ee
we obtain
\bea 
16 \pi^2 R_1(x) & \approx &  
\Delta(x) \int_0^x\!\diff y\,y \left(y - \frac x4 \right) \frac{m^2(x)-m^2(y)}{x-y} \Delta(y) 
-m^2(x) x^2  \Delta^2(x)\nonumber\\
&-&   \int_x^{\infty}\!\diff y\,  m^2(y) [y^2\Delta^2(y)]', \nonumber \\
16 \pi^2 R_2(x) &\approx&  \frac{1}{2} \int_0^x\!\diff y\,y\,m^2(y) \left(1-\frac yx \right) \Delta^2(y) 
+ \frac{1}{4} \int_x^{\infty}\!\diff y\, m^2(y) [y^2\Delta^2(y)]'.
\eea

Finally, since as shown in~\cite{Aguilar:2009nf,Aguilar:2009pp}, $L(x)$ is considerably smaller than $G(x)$ in the entire range of (Euclidean) momenta, we can use the approximation $1+G(x) \approx F^{-1}(x)$; thus, we obtain the approximate equation
\be 
m^2(x) = m^2(0) \frac{F(x)}{F(0)} + \frac{\alpha_s C_A}{2\pi}\,F(x)\, \overline{\!R}(x),  
\label{mass-num-eq}
\ee
with
\bea
\overline{\!R}(x) &=& 
\frac{1}{2} \int_0^x\!\diff y\,y\,m^2(y) \left(1-\frac yx \right) \Delta^2(y) + 
\Delta(x) \int_0^x\!\diff y\,y \left(y - \frac x4 \right) \frac{m^2(x)-m^2(y)}{x-y} \Delta(y)
\nonumber\\
&-&  m^2(x)\, x^2 \Delta^2(x) + \frac{3}{4} \int_0^x\!\diff y\, m^2(y) [y^2\Delta^2(y)]', 
\eea
and $m^2(0)$ given in Eq.~(\ref{m20}).  Evidently, $\overline{\!R}(0) = 0$. 

\subsection{Lattice ingredients: Gluon propagator and ghost dressing function}

The two main ingredients of the mass equation~(\ref{mass-num-eq}) 
are the gluon propagator $\Delta(q^2)$ and the ghost dressing function $F(q^2)$. 
Of course,   
$\Delta(q^2)$ is composed by $J(q^2)$ and $m^2(q^2)$, as dictated by Eq.~(\ref{massive}),  
but, as mentioned in the Introduction, the 
derivation of the corresponding equation for $J(q^2)$ is beyond our powers at this point, mainly due to 
lack of knowledge of certain of its ingredients. Similarly, 
$F(q^2)$ satisfies its own SDE (see, e.g., \cite{Aguilar:2009nf}), 
which would furnish yet another 
equation in a complicated coupled system. For the purposes of the present work, 
which is the preliminary scrutiny of  the mass equation~(\ref{mass-num-eq})  
appearing for the first time in the literature, 
we will instead resort to the high quality lattice data available, 
and use them as inputs inside~(\ref{mass-num-eq}).  

\begin{figure}[!t]
\hspace{-1cm}
\begin{minipage}[b]{0.45\linewidth}
\centering
\includegraphics[scale=0.6]{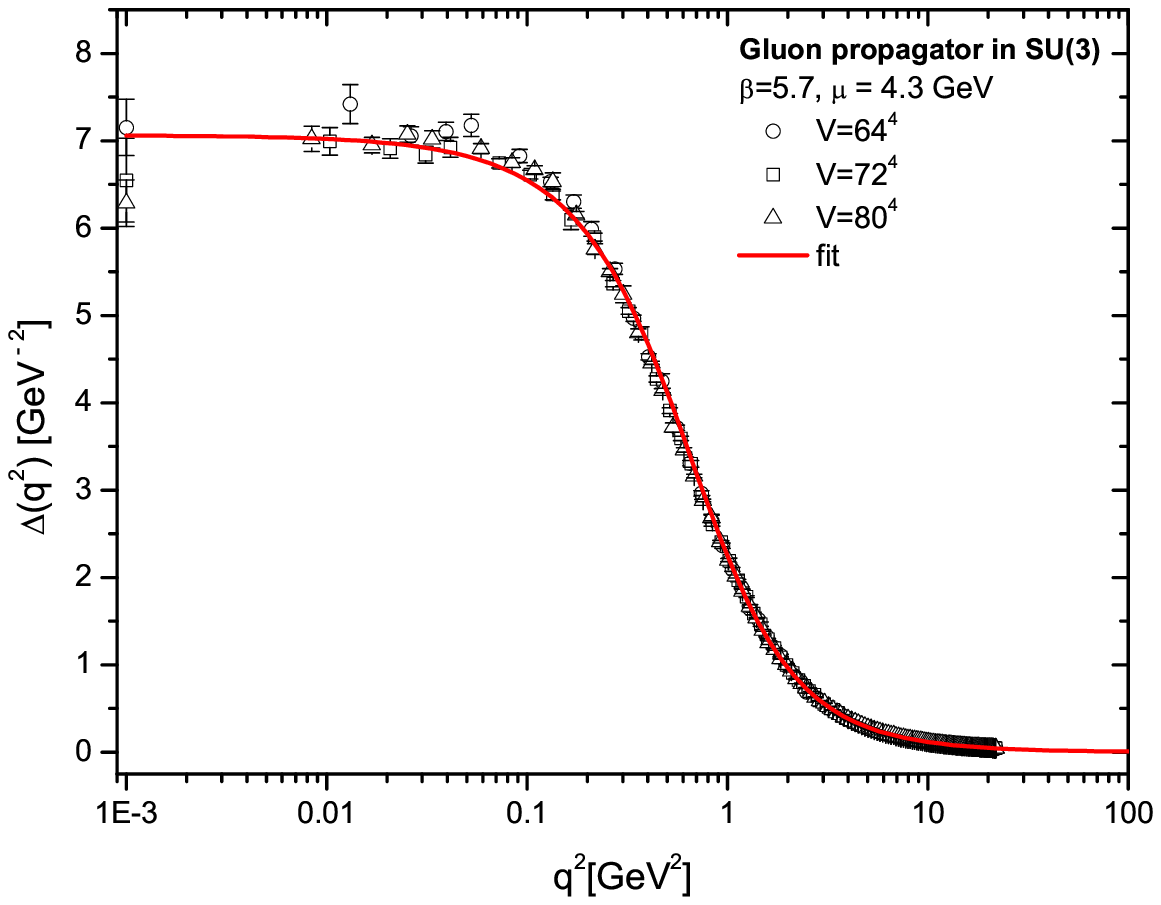}
\end{minipage}
\hspace{0.5cm}
\begin{minipage}[b]{0.50\linewidth}
\includegraphics[scale=0.6]{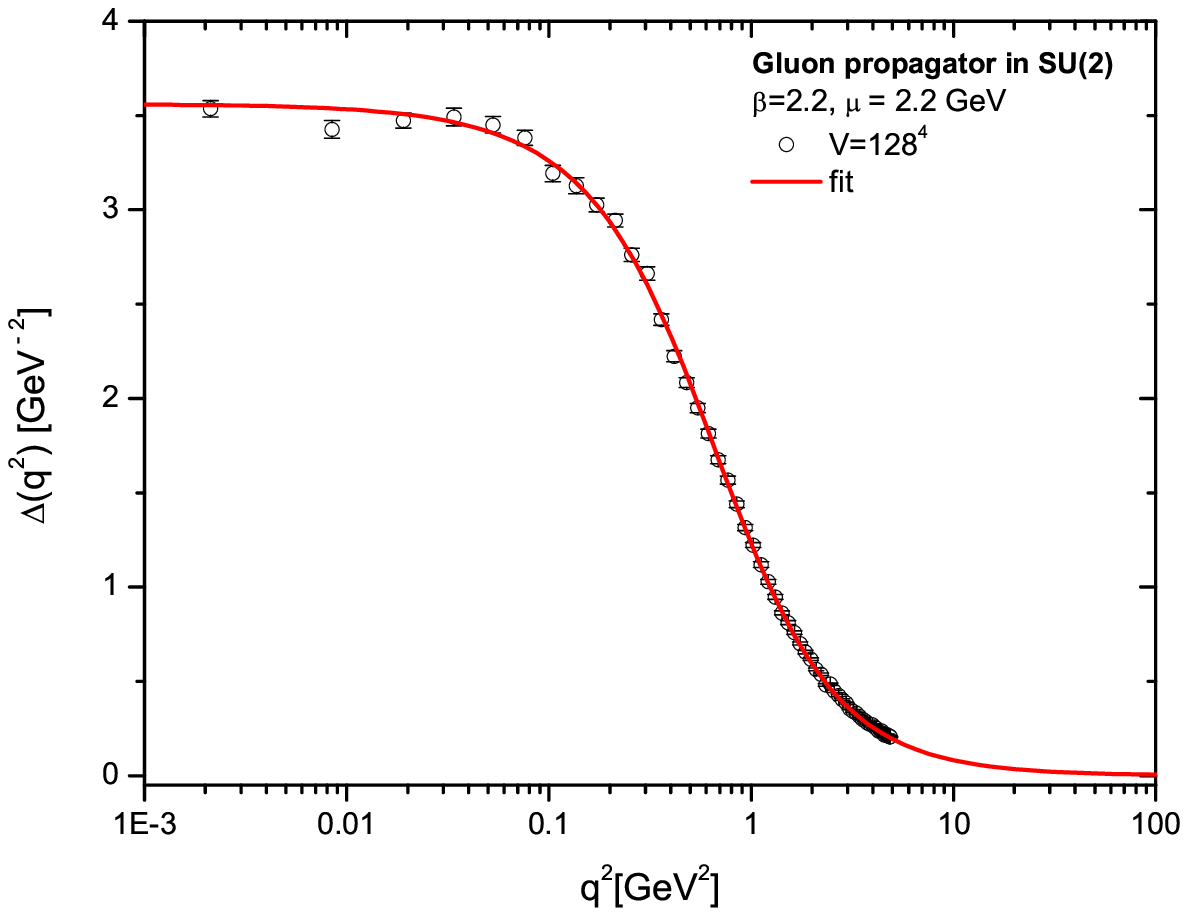}
\end{minipage}
\vspace{-0.5cm}
\caption{Lattice results for the $SU(3)$ (left) and $SU(2)$ (right) gluon propagator, renormalized  at $\mu=4.3$ GeV and $\mu=2.2$ GeV respectively. The continuous lines represents our best fits to the data
obtained from Eq.~(\ref{gluon}).}
\label{gprop}
\end{figure}

In order to do that, 
we start by showing on the left panel of Fig.~\ref{gprop} the lattice data for $\Delta(q^2)$ obtained
in~\cite{Bogolubsky:2007ud}, corresponding to a $SU(3)$ quenched lattice simulation, 
renormalized at \mbox{$\mu=4.3$ GeV};  on the right panel of the same figure, we show 
the quenched $SU(2)$ lattice data obtained in~\cite{Cucchieri:2007md}, 
renormalized at \mbox{$\mu=2.2$ GeV}.

As has been discussed in detail in the literature~\cite{Cornwall:1981zr,Aguilar:2006gr,Aguilar:2008xm}, both sets 
of lattice data can be accurately fitted in 
terms of a IR finite gluon propagator of the form~\cite{Aguilar:2010gm}
\be
\Delta^{-1}(q^2)= M^2(q^2) + q^2\left[1+ \frac{13C_{\rm A}g_1^2}{96\pi^2} 
\ln\left(\frac{q^2 +\rho_1\,M^2(q^2)}{\mu^2}\right)\right],
\label{gluon}
\ee  
where~\cite{Aguilar:2007ie}
\be
M^2(q^2) = \frac{m_0^4}{q^2 + \rho_2 m_0^2}.
\label{dmass}
\ee
The function $M^2(q^2)$ controls the value of  $\Delta^{-1}(q^2)$ at the origin; evidently, 
\mbox{$\Delta^{-1}(0) = M^2(0) = m_0^2/\rho_2$}.
The best fits (shown by the continuous lines in Fig.~\ref{gprop}) correspond 
to the following values of the fitting parameters: 

\begin{itemize}

\item $SU(3)$ case: $m_0= 520$ MeV, $g_1^2=5.68$, $\rho_1=8.55$, $\rho_2=1.91$;  

\item $SU(2)$ case: $m_0= 867$ MeV, 
$g_1^2=10.80$, $\rho_1=1.96$, $\rho_2=2.68$.

\end{itemize} 

\begin{figure}[!t]
\hspace{-1cm}
\begin{minipage}[b]{0.45\linewidth}
\centering
\includegraphics[scale=0.6]{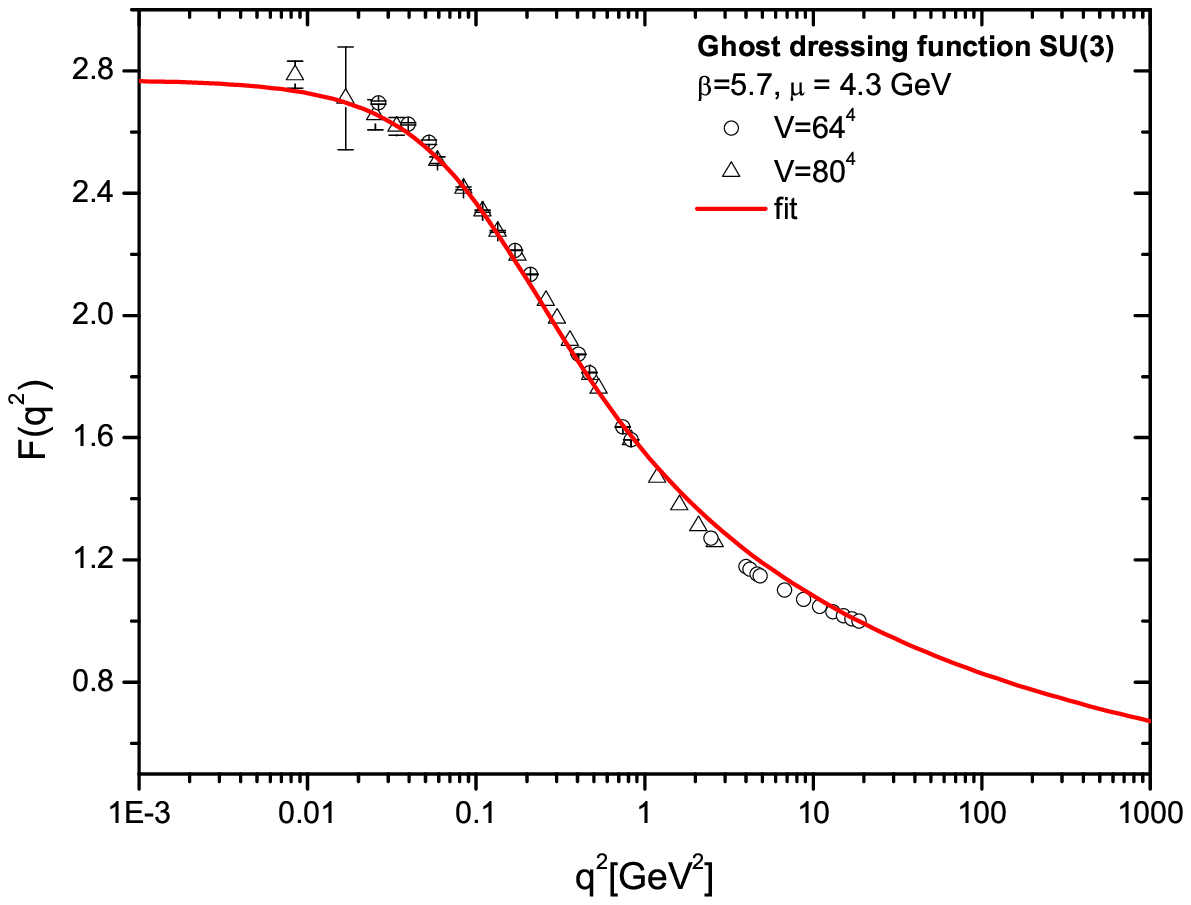}
\end{minipage}
\hspace{0.5cm}
\begin{minipage}[b]{0.50\linewidth}
\includegraphics[scale=0.6]{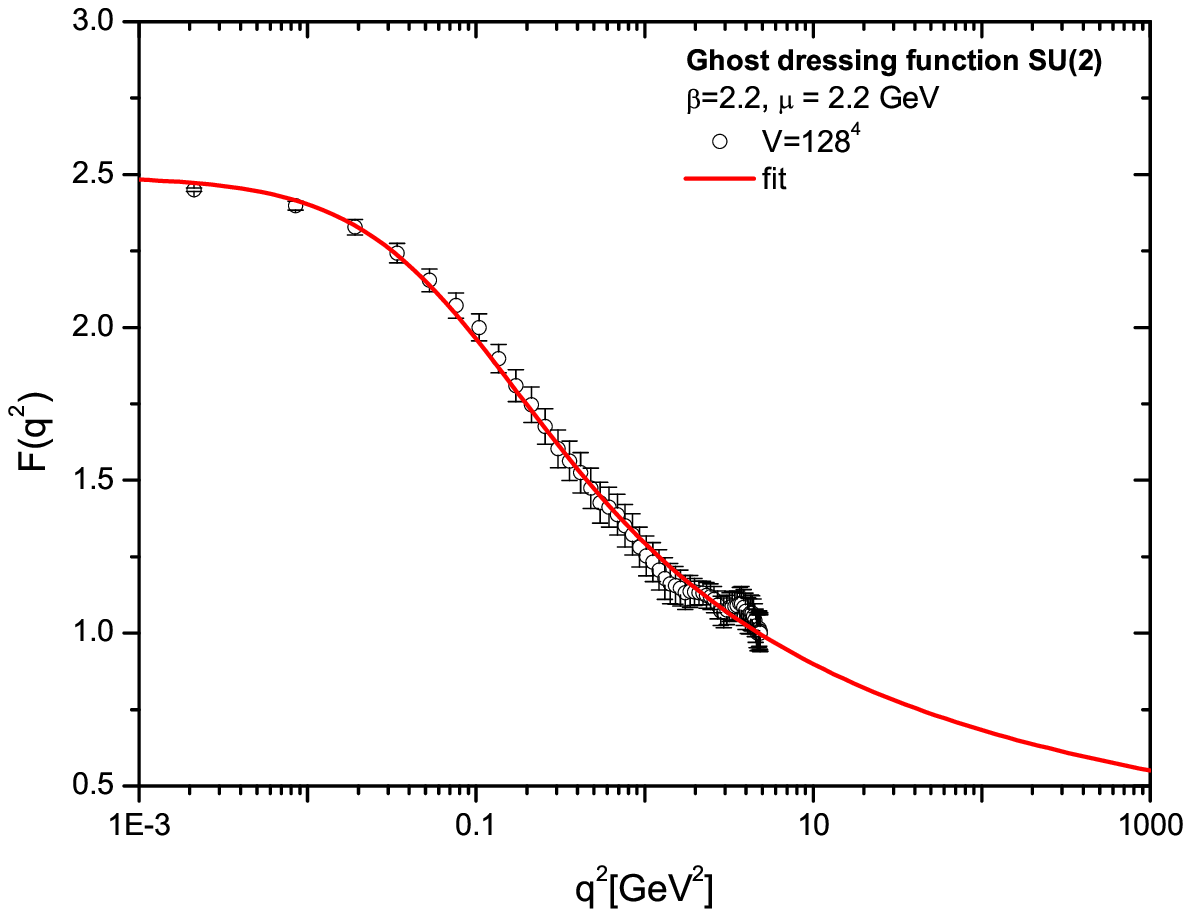}
\end{minipage}
\vspace{-0.5cm}
\caption{Lattice results for the $SU(3)$ (left) and $SU(2)$ (right) ghost dressing function, renormalized at $\mu=4.3$ GeV and $\mu=2.2$ GeV respectively. The continuous lines represent our best fits to the data
obtained from Eq.~(\ref{ghost}).}
\label{ghprop}
\end{figure}

Turning next to the ghost dressing function, on the left panel of Fig.~\ref{ghprop}, 
we show the $SU(3)$ lattice results of~\cite{Bogolubsky:2007ud}, 
renormalized as before at \mbox{$\mu=4.3$ GeV}; on the right panel we plot instead  the 
results for the $SU(2)$ case~\cite{Cucchieri:2007md}, renormalized at \mbox{$\mu=2.2$ GeV}. 
As can be clearly seen, both functions saturate in the deep IR at the 
constant value~\cite{Aguilar:2008xm,Boucaud:2008ji,RodriguezQuintero:2010ss}, and can therefore be fitted in terms of  the expression 
\be
F^{-1}(q^2)= 1+ \frac{9}{4}\frac{C_{\rm A}g_2^2}{48\pi^2}
\ln\left(\frac{q^2 +\rho_3  M^2(q^2)}{\mu^2}\right),
\label{ghost}
\ee
with $M^2(q^2)$ given by Eq.~(\ref{dmass}), but changing 
the parameter $\rho_2 \to \rho_4$. 

The best values for the fitting parameters are:
\begin{itemize}

\item $SU(3)$ case: \mbox{$g_2^2 = 8.57$},
\mbox{$m = 520\,$ MeV}, \mbox{$\rho_3=0.25$}, \mbox{$\rho_4=0.68$};

\item $SU(2)$ case: \mbox{$g_2^2 = 15.03$}, \mbox{$m = 523\,$ MeV}, \mbox{$\rho_3=0.21$}, \mbox{$\rho_4=0.78$}.

\end{itemize}

\subsection{Solutions of the mass equation and extraction of $J_m(q^2)$}

After presenting the precise form of $\Delta(q^2)$ and $F(q^2)$, the next task is to  
find solutions of the approximate mass equation~(\ref{mass-num-eq}).

To begin with, we compute (for both gauge groups considered) the  
derivative of the gluon dressing squared, $[y^2\Delta^2(y)]'$, 
entering into the condition~(\ref{m20}). As mentioned earlier, 
the behavior of this quantity provides a rather direct criterion 
for the existence or not of positive-definite mass solutions, 
and in particular $m^2(0)>0$.
Specifically, the absence of a negative region from this derivative  
immediately excludes such solutions, while   
a relatively shallow ``well'' makes their existence unlikely. 

\begin{figure}[!t]
\hspace{-1cm}
\begin{minipage}[b]{0.45\linewidth}
\centering
\includegraphics[scale=0.6]{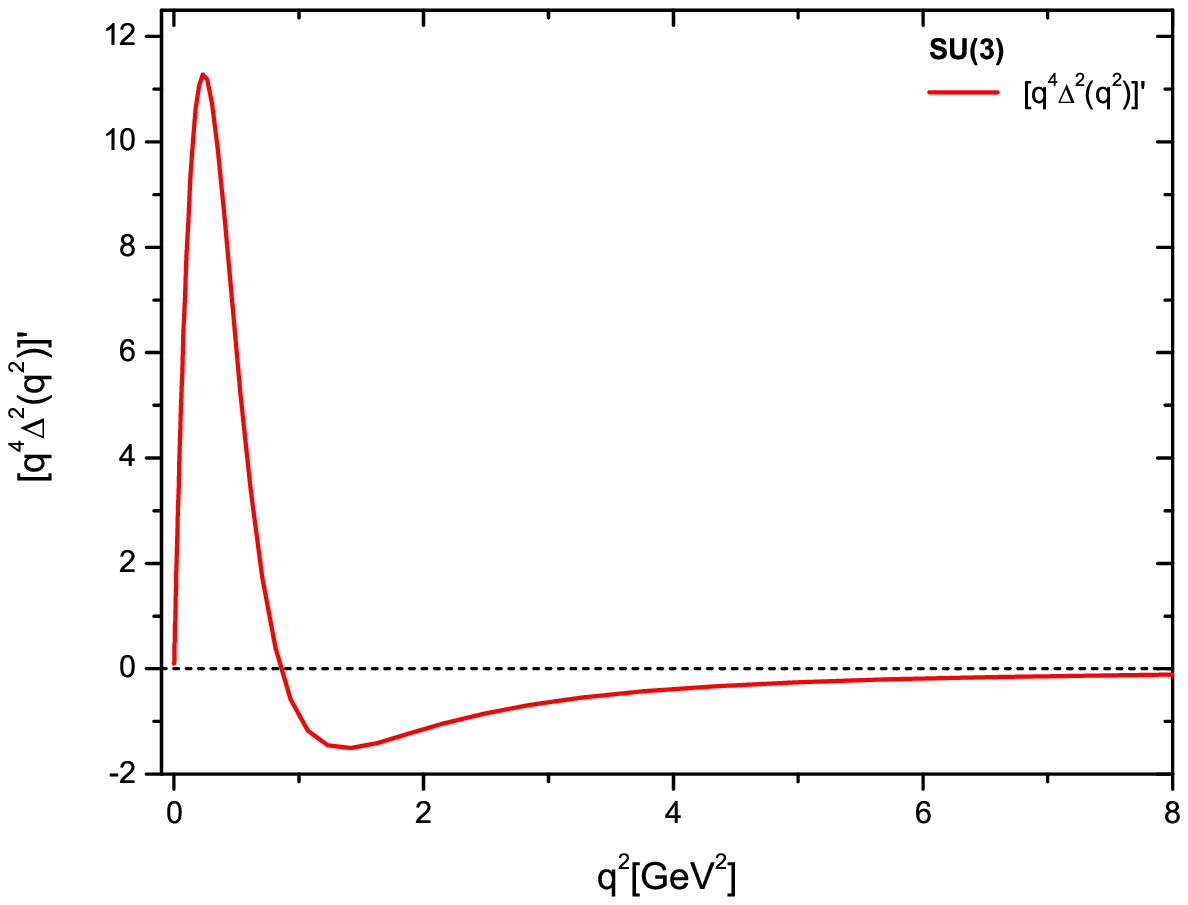}
\end{minipage}
\hspace{0.5cm}
\begin{minipage}[b]{0.50\linewidth}
\includegraphics[scale=0.6]{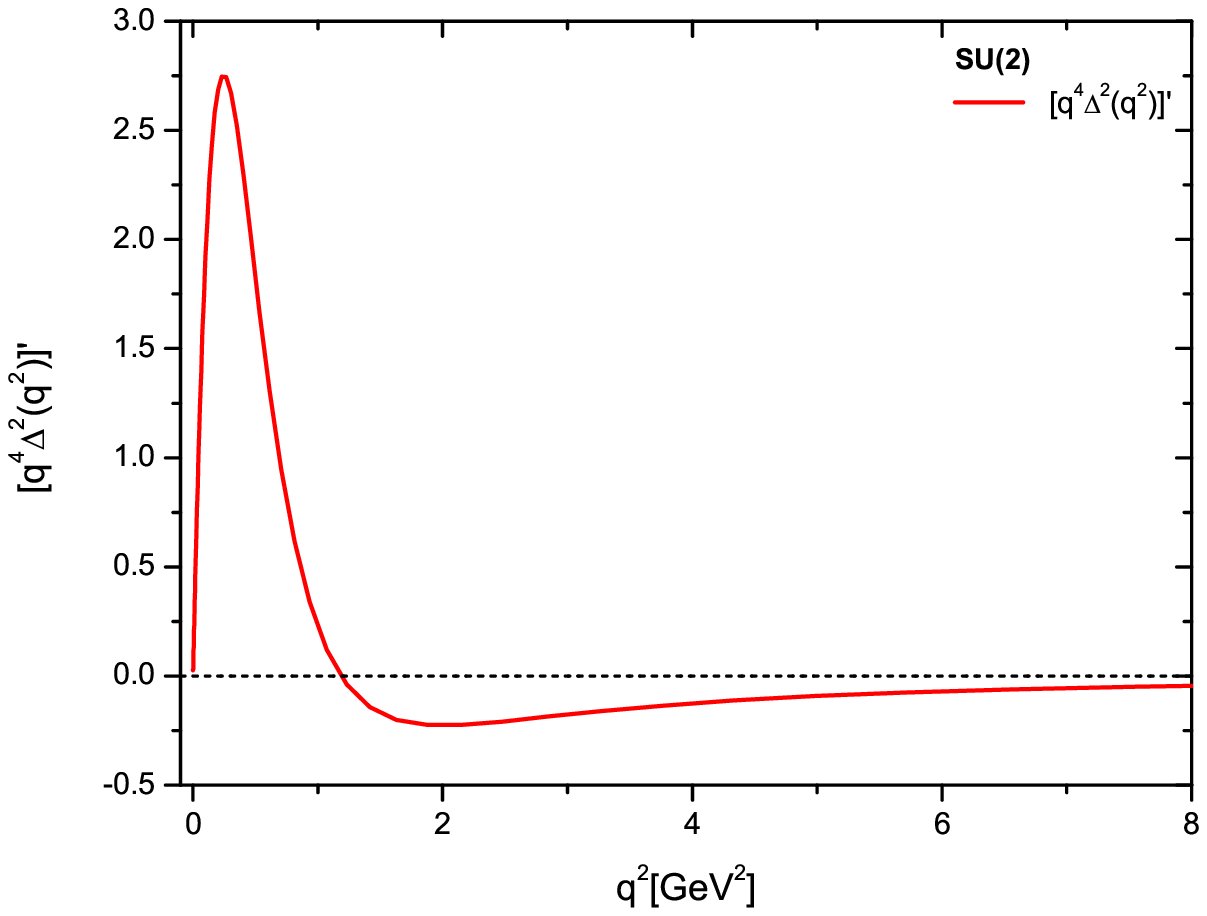}
\end{minipage}
\vspace{-0.5cm}
\caption{The kernel $[q^4\Delta^2(q^2)]'$ appearing in Eq.~(\ref{m20}) 
obtained from the $SU(3)$ (left) and $SU(2)$ (right) lattice data. In both cases one clearly sees the behavior expected for getting a positive value
for $m^2(0)$. The zero crossing happens at $q^2_0\approx0.85$ and $q^2_0\approx 1.1$ respectively.}
\label{kernel4d}
\end{figure}

In the results shown in Fig.~\ref{kernel4d} we clearly 
see that both derivatives change their sign in the 
intermediate momenta region, which, as previously explained, 
constitutes precisely the required behavior. 
This behavior is to be contrasted with that of 
simple propagators, such as $1/(q^2+m^2)$, or 
the Gribov-Zwanziger propagator $q^2/(q^4+m^4)$~\cite{Gribov:1977wm,Zwanziger:1993dh},
which fail to provide the necessary  negative region 
(in fact the derivative is positive everywhere). 
It should be noted that, instead, the ``refined'' version of the Gribov-Zwanziger propagator~\cite{Dudal:2008sp} 
is expected to furnish a considerable negative region, given that it is known to provide a good fit 
to the lattice data. 

Of course, the aforementioned criterion can only serve as a 
necessary but not sufficient condition:
to get a positive definite value for $m^2(0)$ one still needs to demonstrate that 
the negative region $q^2>q^2_0$ (with $q^2_0$ 
the value where the curve is zero) furnishes more support to 
the integral of Eq.~(\ref{m20}) than its positive region. 

To proceed with the actual determination of $m^2(x)$ from Eq.~(\ref{mass-num-eq}), we substitute 
the quantities $\Delta(y)$, $F(y)$ and $C_A$ for the $SU(3)$ and $SU(2)$ gauge groups 
and solve for  the unknown function. 
In both cases the value of $m^2(0)$ is a boundary condition, fixed through 
the value of the corresponding lattice gluon propagator at the origin, {\it i.e.}, $m^2(0)=\Delta^{-1}(0)$. 
Specifically, for $SU(3)$ we have that  $\Delta^{-1}(0)\approx0.14$ while for $SU(2)$
$\Delta^{-1}(0)\approx0.28$

\begin{figure}[!t]
\hspace{-1cm}
\begin{minipage}[b]{0.45\linewidth}
\centering
\includegraphics[scale=0.6]{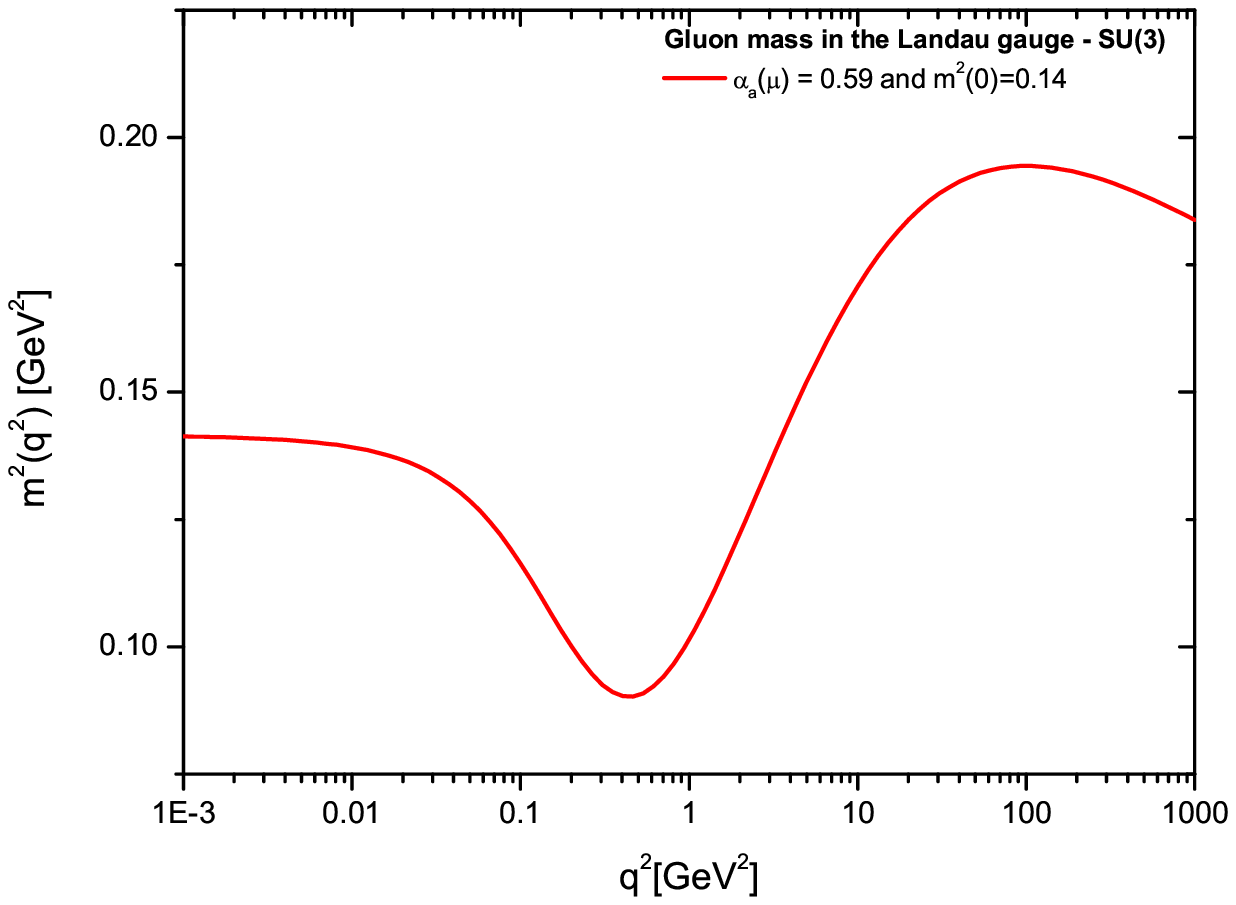}
\end{minipage}
\hspace{0.5cm}
\begin{minipage}[b]{0.50\linewidth}
\includegraphics[scale=0.6]{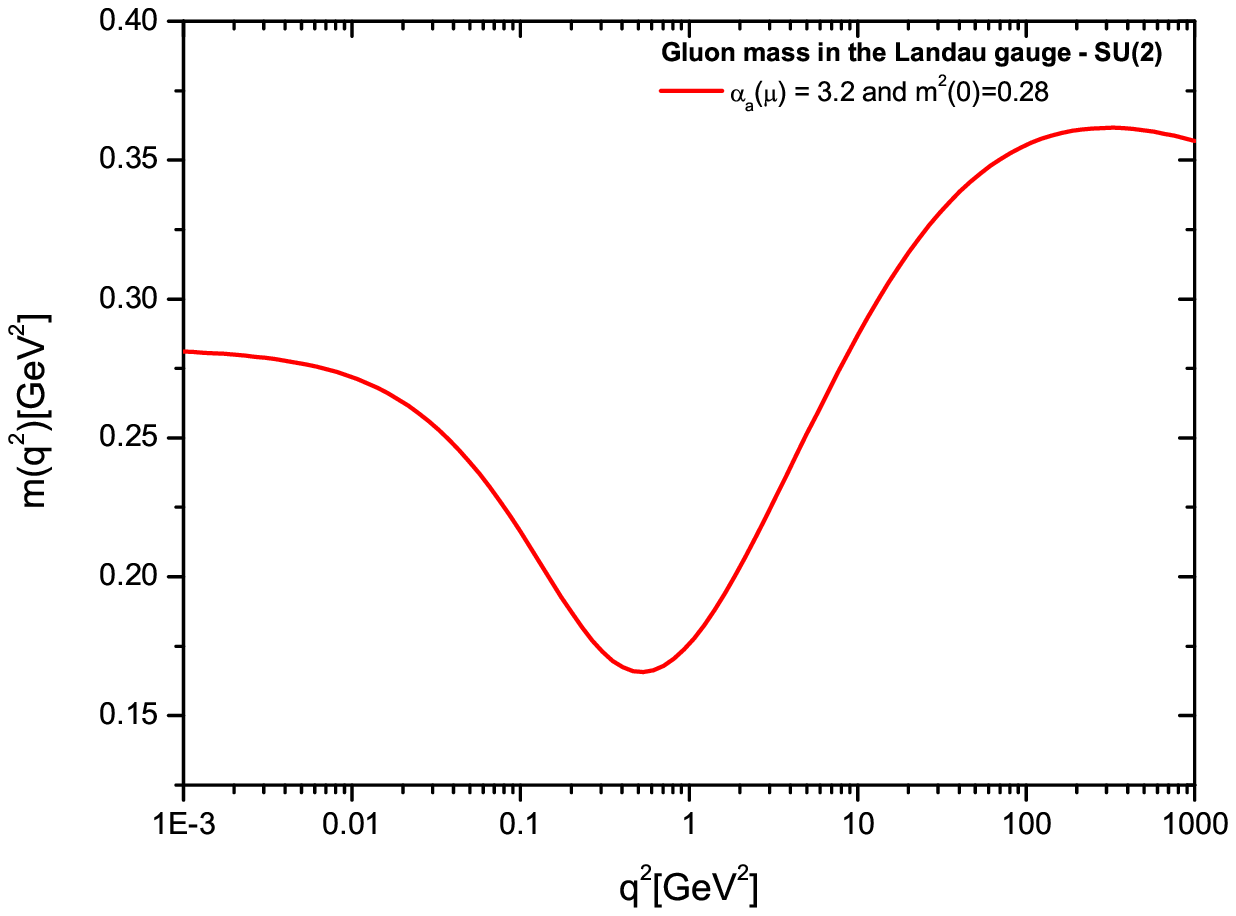}
\end{minipage}
\vspace{-0.5cm}
\caption{The solution for $m^2(q^2)$ obtained through the approximate mass equation~(\ref{mass-num-eq}) 
for $SU(3)$ (left) and $SU(2)$ (right).}
\label{Landau-mass}
\end{figure}

The solutions obtained are shown in Fig.~\ref{Landau-mass}; the values for 
$\alpha_s$ needed to satisfy the boundary condition are $\alpha_s=0.59$ and $\alpha_s=3.2$ for $SU(3)$ and $SU(2)$ 
respectively. Notice that the masses 
corresponding to both gauge groups display the same qualitative behavior, and, as expected, 
are clearly non-monotonic functions of the momentum.

From the solutions for $m^2 (q^2)$ obtained above, and the 
lattice results for $\Delta(q^2)$, we may now extract the approximate 
form of the ``kinetic term'', $J_m (q^2)$. 
Specifically, $J_m(q^2)$ can be determined (in Euclidean space) through Eq.~(\ref{massive}), namely 
\be
J_m (q^2) = \frac{\Delta^{-1}(q^2) -  m^2 (q^2)}{q^2},
\label{jmm}
\ee

\begin{figure}[!t]
\hspace{-1cm}
\begin{minipage}[b]{0.45\linewidth}
\centering
\includegraphics[scale=0.6]{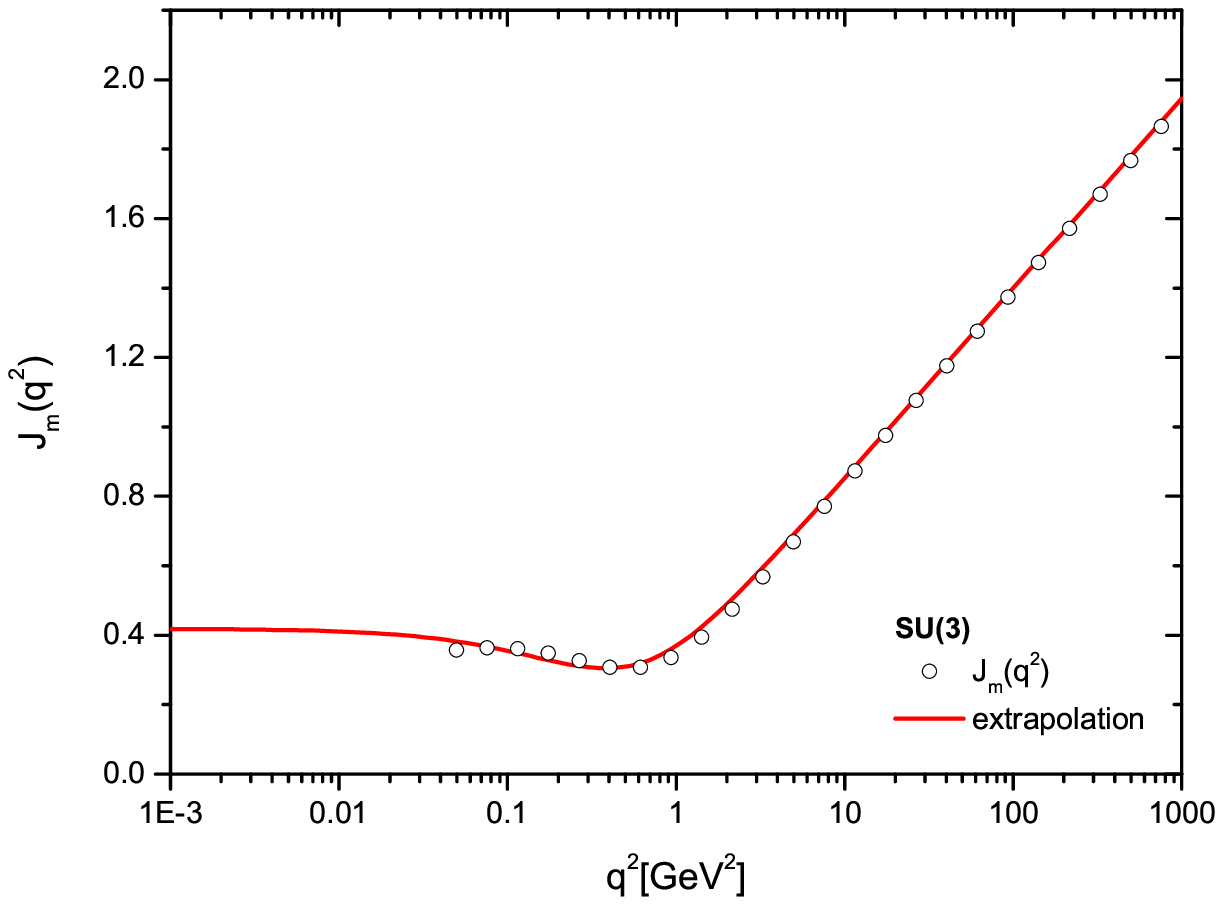}
\end{minipage}
\hspace{0.5cm}
\begin{minipage}[b]{0.50\linewidth}
\includegraphics[scale=0.6]{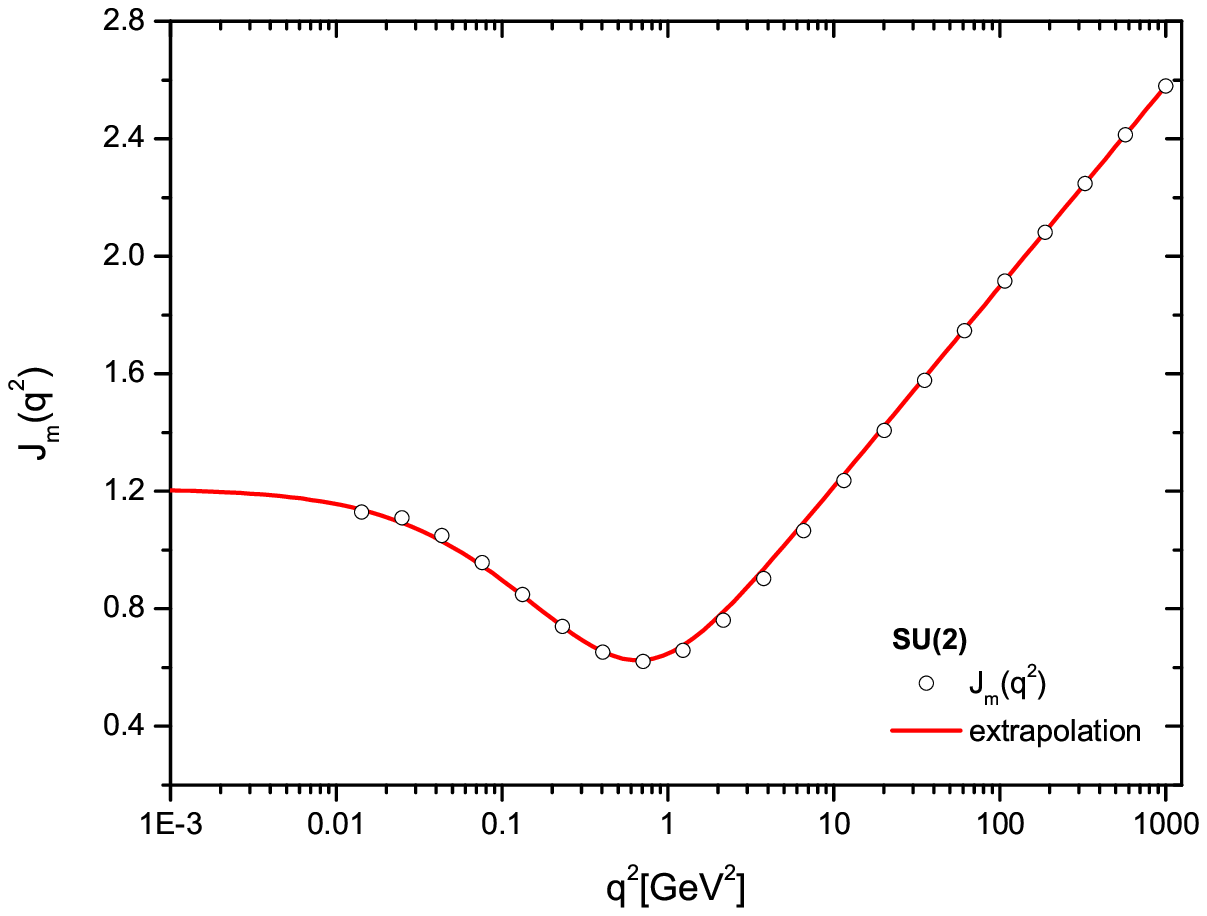}
\end{minipage}
\vspace{-0.5cm}
\caption{Values of  $J_m (q^2)$ obtained from Eq.~(\ref{jmm}) (white circles) using the  $SU(3)$ gluon propagator and the corresponding extrapolation towards the $q^2 \to 0$ limit (continuous line). As usual we show both the $SU(3)$ (left) and the $SU(2)$ cases.}
\label{jm}
\end{figure}

Notice that special care must be taken in the $q^2\to 0$  limit of Eq.~(\ref{jmm}). In 
the region of small momenta, Eq.~(\ref{jmm}) has a delicate cancellation 
between the denominator and the numerator, which also tends to zero in this limit, since $\Delta^{-1}(q^2) \to  m^2 (0)$. 
In order to avoid spurious distortion in the IR behavior 
of  $J_m (q^2)$, we will extract $J_m (q^2)$
until certain (small) value of $q^2$ past which we will do an extrapolation towards  $q^2\to 0$.  
The results of this procedure are shown in Fig.~\ref{jm}, where, 
for both the $SU(3)$ (left) and  the $SU(2)$ (right) cases, we display the 
points obtained directly from Eq.~(\ref{jmm}) as well as our extrapolation curves.

Knowledge of $m^2 (q^2)$ and $J_m (q^2)$ allows one to 
determine the approximate form of the (formally) 
RG-invariant gluon mass that appears 
naturally 
in the definition of the QCD effective charge~\cite{Cornwall:1981zr,Cornwall:1989gv,Binosi:2002ft,Watson:1996fg}.
Let us recall that, 
due to the Abelian WIs satisfied by the PT-BFM Green's functions, the 
propagator $\widehat\Delta(q^2)$ absorbs all the RG logarithms, 
exactly as happens in QED with the photon self-energy.  As a result, the product  
\be
{\overline {\!d\,}}_0(q^2) \equiv g^2_0 \widehat\Delta_0(q^2) = 
g^2 \widehat\Delta(q^2) \equiv {\overline {\!d\,}} (q^2),
\ee
forms a RG-invariant \mbox{($\mu$-independent)} quantity.
As has been explained in the recent literature~\cite{Aguilar:2009ke}, ${\overline {\!d\,}} (q^2)$ may be cast in the form 
\be
{\overline {\!d\,}}(q^2) = \frac{{\overline g}^2(q^2)}{q^2 + \mrgi^2 (q^2) },
\ee
with
\bea
{\overline g}^2 (q^2) &=& g^2 {\widehat J}_m^{-1} (q^2),
\nonumber\\
{\mrgi}^2 (q^2) &=& {\widehat m^2} (q^2) {\widehat J}_m^{-1} (q^2).
\label{RG-terms}
\eea
The two factors defined above are individually RG-invariant; 
the dimensionful quantity corresponds to a massive propagator with a momentum dependent mass, while 
the dimensionless factor ${\overline g}^2 (q^2)/4 \pi$ defines the effective charge.

Next, using the BQIs~(\ref{BQIs}) to relate the components of $\widehat\Delta(q^2)$ to the corresponding ones of $\Delta(q^2)$, we get 
\bea
{\widehat J}_m(q^2) &=& [1+G(q^2)]^2 J_m(q^2),
\nonumber\\ 
{\widehat m}^2 (q^2) &=& [1+G(q^2)]^2 m^2 (q^2),
\eea
and therefore
\be
{\widehat m}^2 (q^2) {\widehat J}_m^{-1} (q^2) = {m}^2 (q^2) {J}_m^{-1} (q^2), 
\ee
which finally furnishes the relation 
\be
{\mrgi}^2 (q^2) = m^2 (q^2) J_m^{-1} (q^2).
\label{rgim}
\ee

\begin{figure}[!t]
\begin{minipage}[b]{0.45\linewidth}
\centering
\includegraphics[scale=0.6]{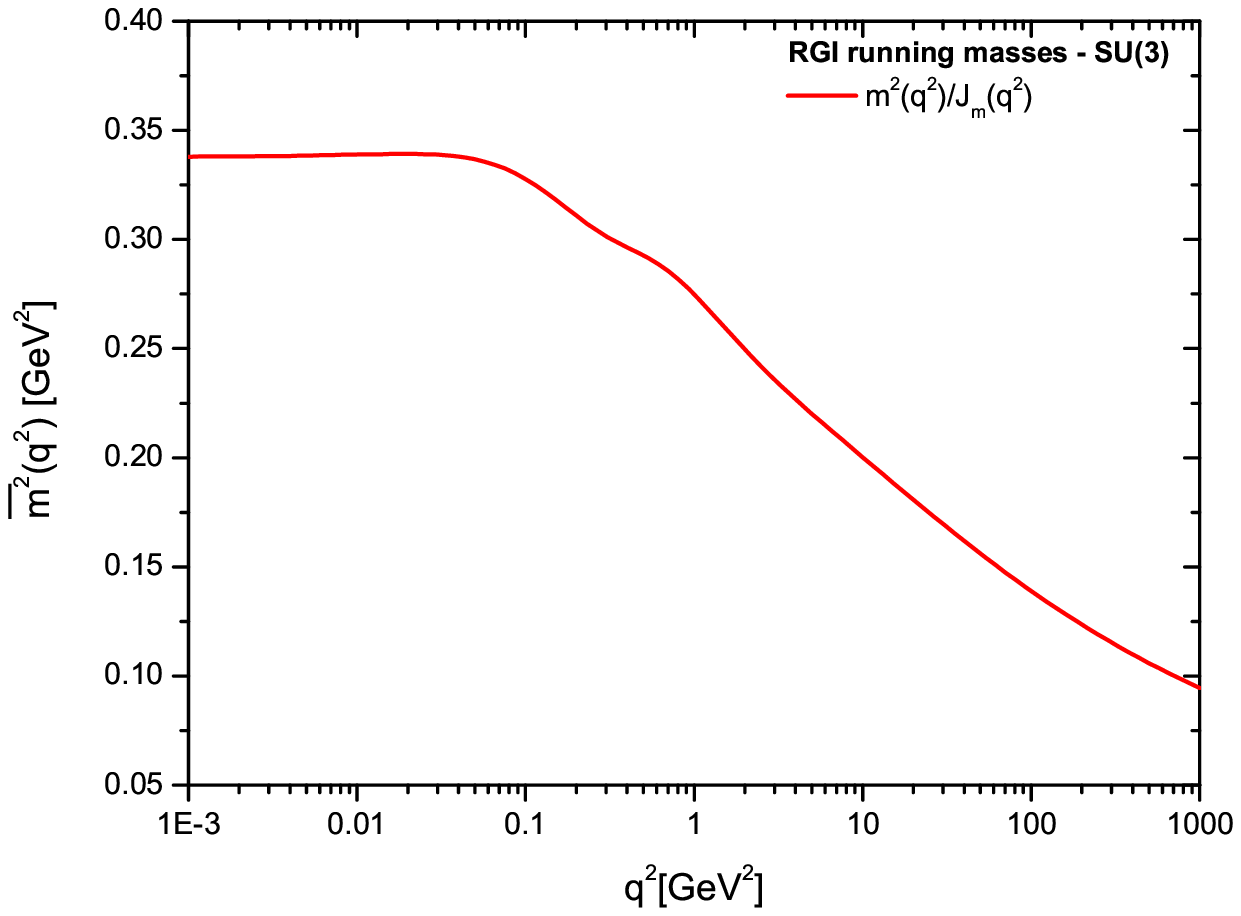}
\end{minipage}
\hspace{0.5cm}
\begin{minipage}[b]{0.50\linewidth}
\includegraphics[scale=0.6]{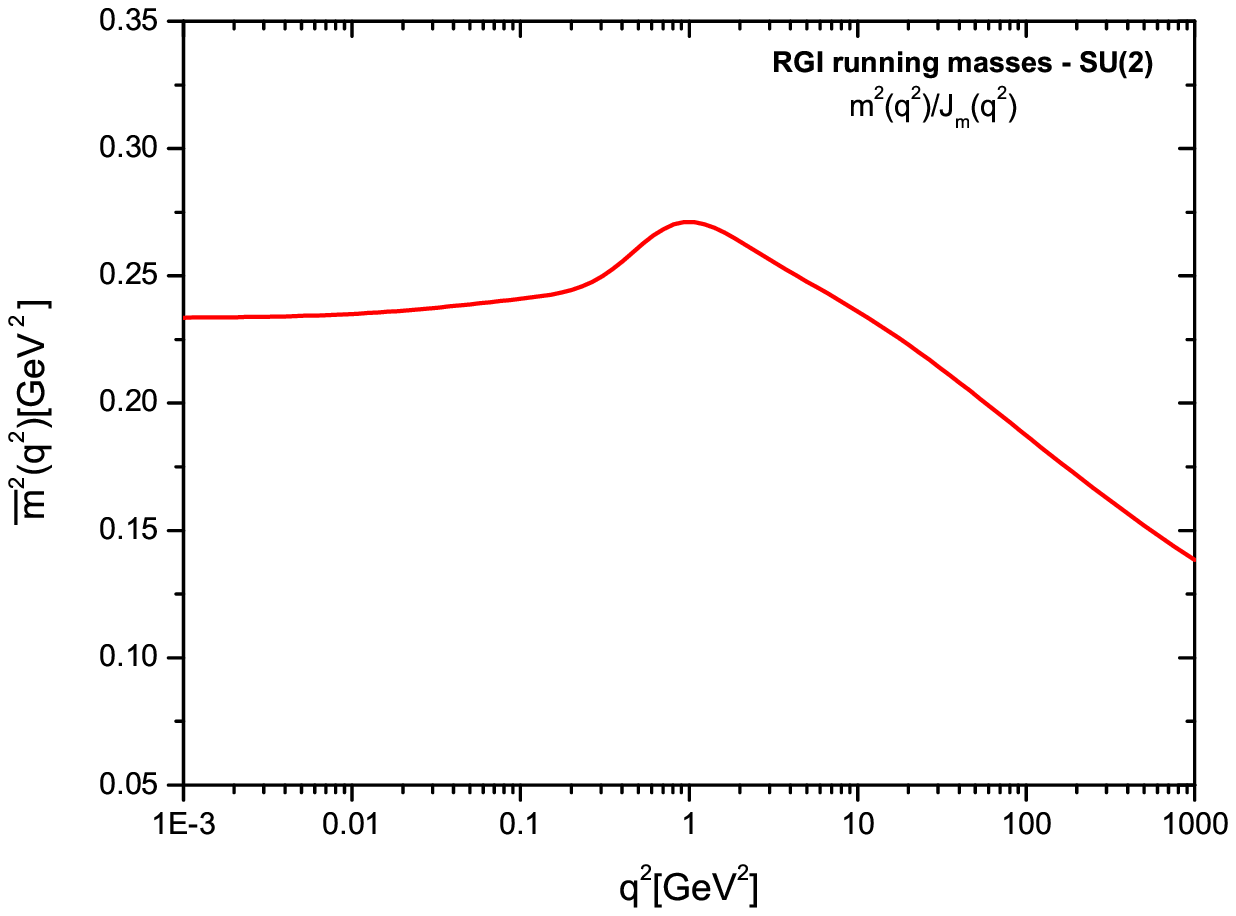}
\end{minipage}
\vspace{-1.0cm}
\caption{The RG-invariant mass, ${\mrgi}^2 (q^2)$, defined in Eq.~(\ref{rgim})
for the $SU(3)$ (left) and $SU(2)$ (right) cases.}
\label{rgi_mass}
\end{figure}

We are now in the position to determine the mass  
${\mrgi}^2 (q^2)$ by simply forming the ratio 
of the plots presented in Fig.~\ref{Landau-mass} and~\ref{jm}.  
The result is  shown in Fig.~\ref{rgi_mass}; as can be seen,
in the $SU(3)$ case ${\mrgi}^2 (q^2)$ corresponds roughly to a monotonically decreasing function 
(see also~\cite{Oliveira:2010xc}),   
with \mbox{${\mrgi} (0)\approx 580 \,\mbox{MeV}$}. Finally, for the $SU(2)$ case
we  obtain \mbox{${\mrgi} (0)\approx 480 \,\mbox{MeV}$}. 

\section{\label{Conc}Discussion and conclusions}

In the present work we have derived the dynamical equation that determines the evolution of the 
gluon mass in the Landau gauge, using as our starting point the ``one-loop dressed'' SDE for the 
gluon propagator in the PT-BFM scheme. 
The entire construction hinges on the crucial assumption that a special vertex, denoted by $V$,  
is dynamically generated, according to the philosophy and formalism 
associated with the Schwinger mechanism. The role of this vertex is to 
maintain gauge invariance (as expressed through the STIs satisfied  
by the Green's functions of the theory) in the presence of a dynamical mass. 
Interestingly enough, the derivation of the mass equation 
does not depend on the specific closed form of that vertex. 

The equation for the gluon mass derived here, given in (\ref{me-final}), 
and in particular its limit in the deep IR, imposes a rather strong constraint 
on the form of the full gluon propagator in the region of intermediate momenta of about
(1-5) ${\rm GeV}^2$.   In this specific range of momenta 
the shape of the gluon propagator must be such that the derivative of the square of the 
gluon dressing function $[q^4\Delta^2(q^2)]'$ becomes 
sufficiently negative, thus ensuring eventually the positivity of the gluon mass.

We emphasize that the central result of this article,  Eq.~(\ref{me-final}), 
does not exhaust all possible contributions to the gluon mass equation. 
Specifically, Eq.~(\ref{me-final}) captures only the 
part of the equation originating from the 
``one-loop dressed'' gluon SDE. In order to determine the 
corresponding contribution coming from the ``two-loop dressed'' gluon SDE
one must identify the seagull cancellation mechanism 
(and the corresponding ``seagull-identity'') that 
operates at the ``two-loop dressed'' level. 
The  identification of the ``two-loop dressed''  
analogue of Eq.~(\ref{seagull})
requires (among other things) some very specific 
information on the structure of the four-gluon vertex, at least in the special kinematic 
limit of vanishing external momentum. Calculations in this direction are already in progress.

It is important to warn the reader about some additional limitations 
afflicting the present work, 
related to the renormalization properties of the mass equation, 
and the dependence of the various quantities, most importantly of the gluon mass, 
on the renormalization point $\mu$.
When dealing with the mass equation of Eq.~(\ref{me-final})
we have tacitly assumed that the 
multiplicative renormalization has been carried out, 
thus rendering all quantities finite (but $\mu$-dependent). In carrying out the SDE renormalization 
one usually resorts to the momentum-subtraction (MOM) scheme;   
in our case this choice is further motivated by the additional fact that this is the scheme employed    
for the renormalization of the lattice data that are used as input into Eq.~(\ref{me-final}).
Of course, given the gluon mass in the Landau gauge is not a RG-invariant quantity, 
there is a residual dependence on $\mu$, which, in principle, should cancel out 
against analogous contributions when a RG-invariant combination is formed (this 
type of powerful cancellation has been presented in~\cite{Aguilar:2009nf} for the QCD effective charge).
However, the approximations employed in the process of the renormalization may distort the 
exact dependence on $\mu$. Specifically, 
the renormalized version of 
Eq.~(\ref{me-final}) displays a 
dependence on some of the renormalization constants $Z$ involved, as happens typically 
in the treatment of SDEs. This fact in itself is normal, but makes the further treatment 
ambiguous, because the correct cancellation of the residual dependence on the UV cutoff 
(induced by the presence of the $Z$)   
requires the knowledge (among other things) of the transverse (automatically conserved) 
part of the full vertex $\bqq$~\cite{Binosi:2011wi}. 
Therefore, the next step has been to set $Z=1$, a fact which, in general, is known 
to alter the dependence of the solution (in this case of $m^2$ ) on $\mu$.   
In fact, the situation appears to be very similar to what happens typically in the 
studies of chiral symmetry breaking through the standard gap equation. 
In this latter context, the various approximations associated with renormalization introduce 
characteristic artifacts; for example, the value of the anomalous dimension 
of the dynamical quark mass is distorted, a problem that is usually compensated by 
modifying accordingly (by hand) the kernel of the gap equation. 
Needless to say, it would be very important to improve on any of the above points, but 
at present this appears to be technically rather difficult.

Given that the existence of a non-trivial vertex $V$ is of central importance, it would be 
absolutely essential to establish its existence. 
This can be done  
following two distinct but complementary approaches. 
First, one may write down the most general longitudinal structure allowed by Lorentz symmetry 
and then use the WI and STIs that the $V$ is supposed to satisfy [{\it e.g.}, (\ref{winp})] 
to actually determine the form of the various form factors, in the spirit of~\cite{Binosi:2011wi}. 
Second, one may address the dynamical question  
of whether such a nonperturbative vertex may be actually produced by the 
strongly coupled Yang-Mills theory. In fact, 
the main characteristic of the vertex $V$, 
which sharply differentiates it from ordinary vertex contribution, 
is that it contains massless bound-state poles. 
In principle, the dynamical formation of such poles must be studied by means of 
a homogeneous Bethe-Salpeter equation, following the methodology 
developed in~\cite{Jackiw:1973ha,Eichten:1974et,Poggio:1974qs}. 
We hope to be able to pursue some of these points in the near future.

\acknowledgments 

The research of J.~P. is supported by the European FEDER and  Spanish MICINN under 
grant FPA2008-02878, and the Fundaci\'on General of the UV. The work of  A.C.A  is supported by the Brazilian
Funding Agency CNPq under the grant 305850/2009-1 and project 474826/2010-4 .

\appendix

\section{\label{app1} Explicit form of the vertex $\bqq$ and $\bqq_m$}

The longitudinal part of the vertex $\bqq$ (and therefore also that of $\bqq_m$)  
has been constructed in~\cite{Binosi:2011wi}  
by simultaneously solving the Ward and Slavnov-Taylor identities presented in Eq.~(\ref{STIs}); for the kinematics, see Fig.~\ref{3g-vertex}.
Specifically, the longitudinal part is written as  
\be
\bqq^{\alpha\mu\nu}(q,r,p)=\sum_{i=1}^{10}X_i(q,r,p)\ell_i^{\alpha\mu\nu}(q,r,p),
\label{tenlon}
\ee
in the standard basis $\ell_i$ of~\cite{Ball:1980ax}
\be
\begin{tabular}{lll}
$\ell_1^{\alpha\mu\nu} =  (q-r)^{\nu} g^{\alpha\mu}$
& 
$\ell_2^{\alpha\mu\nu} =  - p^{\nu} g^{\alpha\mu}$\hspace{.75cm}
&
$\ell_3^{\alpha\mu\nu} =  (q-r)^{\nu}[q^{\mu} r^{\alpha} -  (q\cdot r) g^{\alpha\mu}] $\\
$\ell_4^{\alpha\mu\nu} = (r-p)^{\alpha} g^{\mu\nu}$
&
$\ell_5^{\alpha\mu\nu} =  - q^{\alpha} g^{\mu\nu}$
&
$\ell_6^{\alpha\mu\nu} =  (r-p)^{\alpha}[r^{\nu} p^{\mu} -  (r\cdot p) g^{\mu\nu}]$
\\
$\ell_7^{\alpha\mu\nu} =  (p-q)^{\mu} g^{\alpha\nu}$
&
$\ell_8^{\alpha\mu\nu} = - r^{\mu} g^{\alpha\nu}$
&
$\ell_9^{\alpha\mu\nu} = (p-q)^{\mu}[p^{\alpha} q^{\nu} -  (p\cdot q) g^{\alpha\nu}]$
\\
$\ell_{10}^{\alpha\mu\nu} = q^{\nu}r^{\alpha}p^{\mu} + q^{\mu}r^{\nu}p^{\alpha}$. & &
\end{tabular}
\label{Ls}
\ee
and the $X_i$ are given by
\bea
X_1(q,r,p) &=&  \frac{1}{4}{\widetilde J}(q^2) \left\{ 
- p^2 b_{prq} F(r^2) + 
[2 a_{rpq} + p^2 b_{rpq} + 2 (q\cdot r) d_{rpq} ]F(p^2) \right\}    
\nonumber\\
&+& \frac{1}{4} J(r^2)\left[ 2 +  (r^2-q^2) {\widetilde b}_{qpr} F(p^2)\right]
+ \frac{1}{4} J(p^2)\, p^2 \,{\widetilde b}_{qrp} F(r^2) 
\nonumber\\
X_2(q,r,p)  &=&  \frac{1}{4}{\widetilde J}(q^2)\left\{
(q^2- r^2) b_{prq} F(r^2) + 
[2 a_{rpq} + (r^2-q^2)b_{rpq} + 2 (q\cdot r) d_{rpq} ] F(p^2)\right\}  
\nonumber\\
&+& \frac{1}{4} J(r^2)\left[ - 2 +  p^2 {\widetilde b}_{qpr} F(p^2)\right]
+ \frac{1}{4} J(p^2)\, (r^2- q^2)\,{\widetilde b}_{qrp} F(r^2) 
\nonumber\\
X_3(q,r,p) &=& \frac{F(p^2)}{q^2-r^2} 
\left\{{\widetilde J}(q^2)\left[a_{rpq} - (q\cdot p) d_{rpq} \right] 
-  J(r^2) \left[{\widetilde a}_{qpr} - (r\cdot p){\widetilde d}_{qpr}\right] \right\}
\nonumber\\
X_4(q,r,p) &=& \frac{1}{4}{\widetilde J}(q^2) q^2 \left[b_{prq} F(r^2) + b_{rpq} F(p^2)\right]
+ \frac{1}{4} J(r^2)  \left[2 - q^2 {\widetilde b}_{qpr} F(p^2)\right]
\nonumber\\
&+& \frac{1}{4} J(p^2) \left[2 - q^2 {\widetilde b}_{qrp} F(r^2)\right]
\nonumber\\
X_5(q,r,p) &=& \frac{1}{4}{\widetilde J}(q^2) (p^2-r^2) \left[b_{prq} F(r^2) + b_{rpq} F(p^2)\right]
+ \frac{1}{4} J(r^2)  \left[2 +(r^2-p^2) {\widetilde b}_{qpr} F(p^2)\right]
\nonumber\\
&-& \frac{1}{4} J(p^2) \left[2 +(p^2-r^2) {\widetilde b}_{qrp} F(r^2)\right]
\nonumber\\
X_6(q,r,p) &=& \frac{J(r^2)-J(p^2)}{r^2-p^2}
\nonumber\\
X_7(q,r,p) &=& X_1(q,p,r)
\nonumber\\
X_8(q,r,p) &=& - X_2(q,p,r)
\nonumber\\
X_9(q,r,p) &=& X_3(q,p,r)
\nonumber\\
X_{10}(q,r,p) &=& \frac{1}{2}\left\{ 
{\widetilde J}(q^2) \left[b_{prq} F(r^2) - b_{rpq} F(p)\right]
+   J(r^2) F(p^2) {\widetilde b}_{qpr} - J(p^2) F(r^2) {\widetilde b}_{qrp} \right\}.
\label{X10}
\eea
The functions  $a_{qrp}\equiv a(q,r,p)$, etc are the form factors appearing in the tensorial decomposition of 
the ghost-gluon kernels $H_{\nu\mu}(p,r,q)$ and $\widetilde{H}_{\nu\mu}(p,r,q)$, namely 
\be
H_{\nu\mu}(p,r,q)=g_{\mu\nu}a_{qrp}-r_\mu q_\nu b_{qrp}+q_\mu p_\nu c_{qrp}+q_\nu p_\mu d_{qrp}+p_\mu
p_\nu e_{qrp}, 
\label{abcde}
\ee
and similarly for $\widetilde{H}$.
They satisfy the non-trivial all-order 
constraints 
\bea
& & F(r^2) [{a}_{prq} - (r\cdot p){b}_{prq} + (q\cdot p){d}_{prq}]
= F(p^2)[{a}_{rpq} - (r\cdot p){b}_{rpq} + (q\cdot r){d}_{rpq}],\nonumber \\
& &F(r^2) [{\widetilde a}_{qrp} - (q\cdot r){\widetilde b}_{qrp} + (q\cdot p){\widetilde d}_{qrp}] = 1.
\label{constr}
\eea

\section{\label{app2}On the relation between ${\widetilde m^2}(q^2)$ and $m^2(q^2)$}

In Section~\ref{gml} we have assumed that the relation~(\ref{mtilde-m-1}) between the masses ${\widetilde m}(q)$ and $m(q)$ holds. This is tantamount to claiming that the BQIs~(\ref{BQIs}) hold  after dynamical mass generation has taken place. 

To further substantiate this claim, let us consider the SDE for the $QB$ gluon self-energy $\widetilde{\Pi}$. If we keep dressed the background side of the equation, we can still truncate meaningfully the SDE retaining only the one-loop dressed gluon contributions, which now read
\bea
(b_1)_{\mu\nu}&=& \frac12\,g^2C_A
\int_k\!\Gamma^{(0)}_{\mu\alpha\beta}(q,k,-k-q)\Delta^{\alpha\rho}(k)
\Delta^{\beta\sigma}(k+q)\gfullb_{\nu\rho\sigma}(q,k,-k-q),\nonumber \\
(b_2)_{\mu\nu}&=&g^2C_A \left[g_{\mu\nu}\int_k\!\Delta^\rho_\rho(k)
-\int_k\!\Delta_{\mu\nu}(k)\right].
\label{gl-1ldr-I}
\eea
The projection to the Landau gauge gives rise to 
three terms only, which coincide with $A_1$, $A_2$ and $A_4$ of Eq.~(\ref{thebees}). 
Then writing
\be
\widetilde{\Delta}^{-1}(q^2) \equiv q^2 \widetilde{J}(q^2) - \widetilde{m}^2(q^2) = q^2+i\widetilde{\Pi}(q^2),
\ee
it is relatively straightforward to establish that  
the rhs of the equation for $\widetilde{m}$ is determined by the 
the mass term of $A_1$ only. Specifically, 
using the result~(\ref{A1m}) one has (Euclidean space)
\bea
\widetilde{m}^2(q^2)&=&2g^2C_A\int_k\!\left[k^2-\frac{(k\cdot q)^2}{q^2}\right]\frac{m^2(k+q)-m^2(k)}{(k+q)^2-k^2}
\Delta_m(k)\Delta_m(k+q)\nonumber \\
&=&\frac{g^2C_A}{d-1}\left[A_1+A_2+A_4\right]_{m^2},
\label{mtilde-m-II}
\eea
where in the last step we have used Eq.~(\ref{A1m}). 


Substituting the above result, together with~(\ref{A3mtilde}), into Eq.~(\ref{me-gen}), we find 
\bea
m^2(q^2) &=& \frac{\widetilde m^2 (q^2)}{[1+G(q^2)]^2}+ 
\frac{\widetilde m^2 (q^2)G(q^2)}{[1+G(q^2)]^2}
\nonumber\\
&=& \frac{\widetilde m^2 (q^2)}{1+G(q^2)}\,,
\eea
namely the relation of Eq.~(\ref{mtilde-m-1}).

\end{document}